\documentclass{article}

\usepackage{arxiv}

\usepackage[utf8]{inputenc} 
\usepackage[T1]{fontenc}    
\usepackage{url}            
\usepackage{booktabs}       
\usepackage{amsfonts}       
\usepackage{nicefrac}       
\usepackage{microtype}      
\usepackage{xcolor}

\usepackage{graphicx}
\usepackage{gensymb}
\usepackage{float}

\title{Solid-electrolyte interphases (SEI) in nonaqueous aluminum-ion batteries}

\author{
  Nicolò Canever, Thomas Nann\thanks{Corresponding author}\\
  School of Mathematical and Physical Sciences\\
  The University of Newcastle\\
  Callaghan, NSW 2308\\
  Australia\\
  e-mail: {\texttt{\color{blue}thomas.nann@newcastle.edu.au}}\\
  \And
  Fraser R.\ Hughson\\
  School of Physical and Chemical Sciences\\
  Victoria University of Wellington\\
  Kelburn, Wellington 6012\\
  New Zealand\\
}

\begin{document}

\maketitle

\begin{abstract}
Nonaqueous aluminum-ion batteries are an interesting emerging energy storage technology, offering a plethora of advantages over existing grid energy storage solutions. Carbonaceous and graphitic materials are an appealing cathode material in this system, thanks to their low cost and excellent rate capabilities. The phenomenon of poor Coulombic efficiency in the first cycle, however, is a known issue among some types of carbons, the reasons for which are yet to be fully understood. In this work, we propose that such processes are caused by the formation of a solid-electrolyte interphase, in a similar fashion to graphite anodes in lithium-ion batteries. Using electrospun carbon nanofibers as a model material with tunable crystallinity, the cause of such phenomena was found to be linked to the presence of surface defects in the cathode material, and was further amplified by high surface area. The simple use of a binder polymer, however, helps mitigating the issue by shielding surface defects from direct contact with the electrolyte.
\end{abstract}


\section{Introduction}
The demand for grid-level energy storage is currently on the rise, due to the gradual global transition to a renewable energy economy. Electrochemical batteries are an enabling technology in this shift, thanks to their high versatility and dependability. Current commercially available technologies such as lithium-ion batteries, however, are often lacking desirable features for large-scale applications such as cost-effectiveness and safety. Nonaqueous aluminum-ion batteries (AIBs) are an emerging battery technology showing many advantageous features for large-scale grid-level energy storage. AIBs use non-flammable components derived from earth-abundant, inexpensive materials.\cite{ambroz_trends_2017,canever_acetamide:_2018} Furthermore, the electrochemistry of these systems is based on the reversible plating and stripping of metallic aluminum from a molten salt electrolyte composed of a Lewis-acidic mixture of 1-ethyl-3-methylimidazolium chloride ([EMIm]Cl) and aluminum trichloride (AlCl$_3$) at the anode. This three-electron redox mechanism, combined with the apparent lack, or suppression, of dendrite growth,\cite{chen_oxide_2017} ensures a high theoretical capacity (up to 2980 mAh g$^{-1}$), opening up the possibility of devices with high energy density. Perhaps one of the most outstanding features of AIBs is their ability to withstand supercapacitor-like cycling rates with excellent capacity retention and coulombic efficiencies when graphitic or graphene-based cathodes are employed.\cite{kravchyk_efficient_2017,das_graphene:_2018} These characteristics are also attractive for grid-level energy storage applications, as they could accommodate the rapid surges in power demand typical of these systems.

One common feature of AIBs that use graphitic cathodes is the poor charge/discharge efficiency affecting the devices in first few cycles. This phenomenon has been observed and reported for a variety of both natural and synthetic graphite materials,\cite{kravchyk_efficient_2017,wang_kish_2017,antonioelia_insights_2017,greco_influence_2018} but no detailed description or convincing explanation for this behavior has been given in any of the previous publications. In this work, we propose that these efficiency losses are related to the decomposition of the ionic species in the electrolyte, resulting in the formation of a solid-electrolyte interphase (SEI). A similar process is known to take place in lithium-ion batteries, where graphitic materials are used as the active material at the anode.\cite{gauthier_electrodeelectrolyte_2015} For AIBs, the formation of an SEI film has been briefly mentioned in other reports on graphitic cathodes,\cite{li_prelithiation_2018,wang_novel_2018} however, not enough in-depth studies have been performed to prove this mechanism. The composition of the SEI layer and the mechanism behind its formation were analyzed using both commercially available pyrolytic graphite paper (PGP) and electrospun carbon nanofibers (CNF). Thanks to their tunable degree of graphitic character,\cite{inagaki_carbon_2012} the latter material was particularly useful in determining the underlying causes behind the process: the main factor contributing to the formation of the SEI was found to be the presence of surface defects on the cathode, which are present in most carbonaceous materials, and especially prominent in those with low graphitic character. Such defects can act as “catalytic centers”, promoting the decomposition of the electrolyte species at lower potentials compared with ideal, defect-free graphite. The surface area was also found to be a factor contributing to the process, as materials with a high surface area can provide a higher population of surface defects, amplifying the formation of an SEI. On the other hand, applying a thin coating layer of an inert polymer such as polyvinylidene difluoride (PVDF) onto the cathode has shown to mitigate the process by effectively shielding the surface of the electrode from direct contact with the electrolyte and inhibiting undesired side reactions.

\section{Results and Discussion}

\begin{figure}
  \centering
 \includegraphics[width=0.4\textwidth]{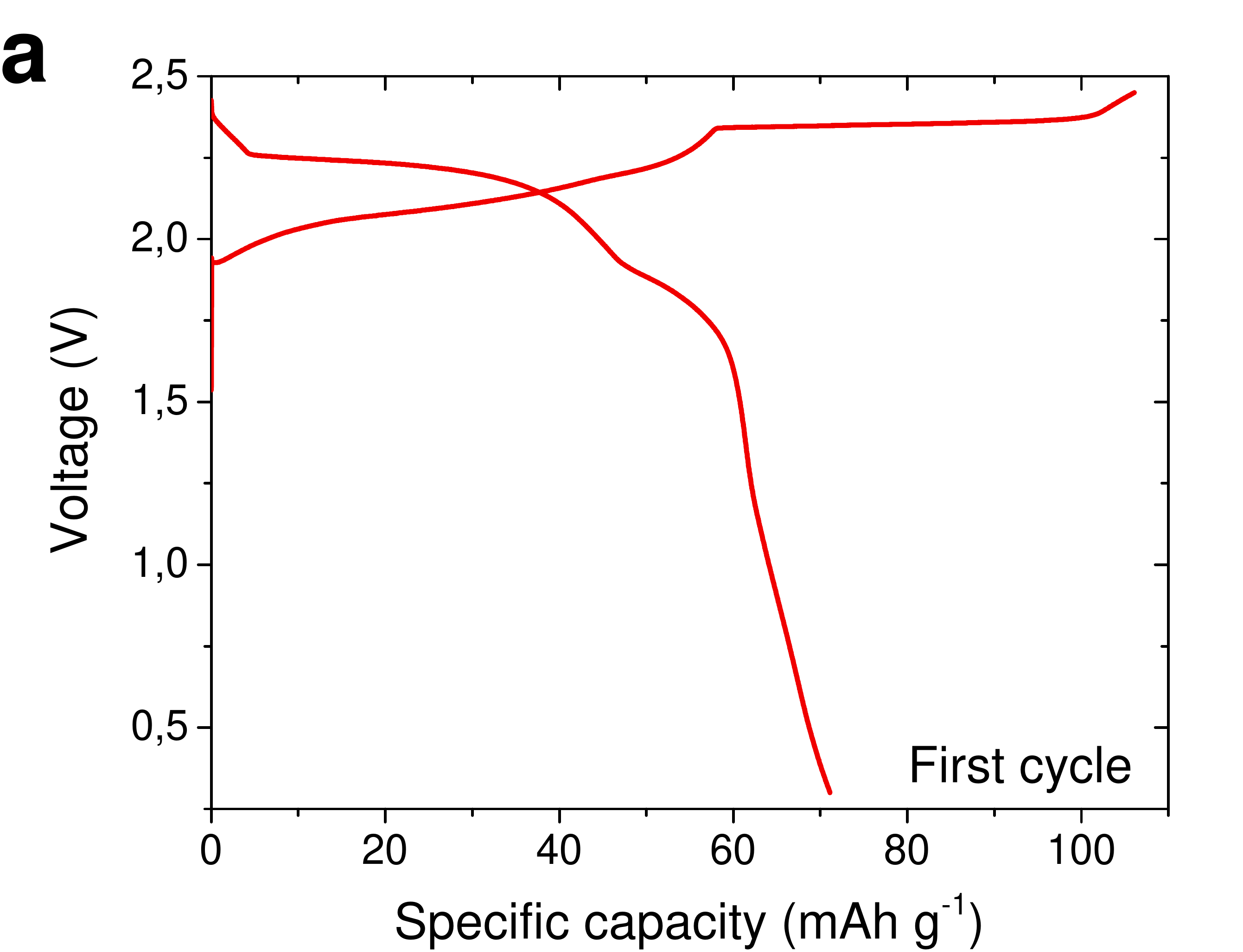}
 \includegraphics[width=0.4\textwidth]{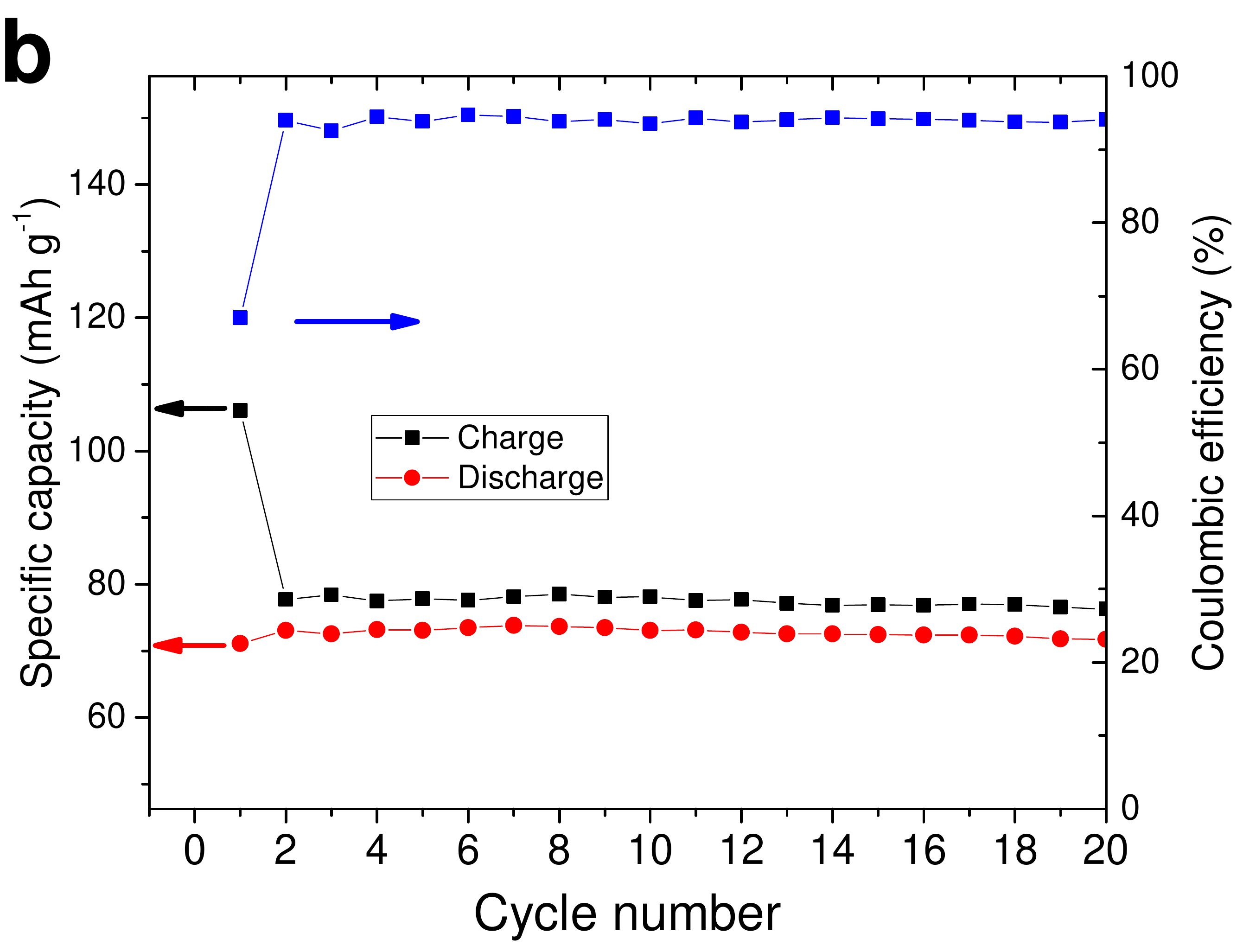}
  \caption{(a) Galvanostatic profiles relative to the first cycle (50 mA g$^{-1}$) of a Swagelok-type cell built using pyrolytic graphite paper. (b) Specific capacities and coulombic efficiencies of the same galvanostatic charge-discharge test.}
  \label{fig:fig1}
\end{figure}

A preliminary galvanostatic charge-discharge test was performed in a Swagelok-type cell using PGP as cathode, at a current rate of 50 mA g$^{-1}$, within the voltage window of 0.3--2.45 V. It can be seen from Figure \ref{fig:fig1} that the first cycle was affected by low coulombic efficiency, as the charging step yielded a specific capacity of 106 mAh g$^{-1}$, compared to the 71 mAh g$^{-1}$ of the corresponding discharge step. In the following cycles, the gap between charge and discharge capacities was reduced considerably, leading to higher (>90\%) efficiencies. This behavior is equivalent to what has been reported by previous reports using pyrolytic graphites. These reports have attributed the phenomenon to a decrease of the porosity of the electrode, caused by the volumetric expansion of the graphitic lattice in the first cycle.\cite{antonioelia_insights_2017,greco_influence_2018} Although this explanation was well argued, we hypothesize that other factors  are also contributing to the observed behavior. Ex-situ scanning electron microscopy (SEM) images (Figure S1
) of the cathode revealed the presence of a thick, lumpy film on the surface, primarily composed of alumnum and chlorine species, as revealed by energy-dispersive X-ray spectroscopy (EDXS). These films could not be removed from the cathode by washing with a variety of protic, polar, and non-polar solvents (Figure 
; it is therefore unlikely that this film was merely composed of the [EMIm]Cl/AlCl$_3$ electrolyte, instead, it is likely  caused by the decomposition of the species making up the ionic liquid.

We  mentioned earlier that the formation of an SEI has been confirmed to exist for graphitic materials when used as anodes for lithium-ion batteries. In these systems, the phenomenon takes place during the first charging of the device, when Li ions intercalate into the graphite anode.\cite{gauthier_electrodeelectrolyte_2015} In the AIB system, on the other hand, carbon constitutes the positive electrode. The intercalation mechanism follows an opposite pattern, as the negatively charged tetrachloroaluminate-ions (AlCl$_4^-$) are inserted into the cathode during the charging phase. Therefore, it is not unreasonable to speculate that the formation of an SEI on graphite, caused by the decomposition of the electrolyte, would primarily happen during the first intercalation event, corresponding to the first charging phase in both battery chemistries, regardless of the polarity of the carbonaceous electrode. For AIBs, this assumption was also supported by a voltage overlap between the intercalation reaction and the oxidation voltage limit of the electrolyte (Figure S2
), suggesting that the decomposition of the electrolyte is in part taking place at the cathode during the charging phase, even at potentials lower than the conventional cut-off value of 2.45 V. In fact, this cut-off voltage was determined empirically in one of the earlier AIB publications by Lin \textit{et al.} \cite{lin_ultrafast_2015} for a “graphitic foam” cathode, and has been since adopted as the upper voltage limit in most publications using graphitic cathodes. The assumptions made for graphitic foam, however, might not be valid for other types of carbonaceous materials, due to the difference in their physical and chemical properties, which could impact the potentials in which the intercalation and decomposition reactions take place. 

Some evidence related to this issue can be found in previous literature. Two recent articles have investigated the performance of zeolite-templated carbon \cite{stadie_zeolite-templated_2017} and acetylene black,\cite{wang_kish_2017} two carbonaceous materials with high specific surface area, as cathodes for AIBs. Although the authors claim that no SEI formation was observed, the electrochemical data reported for these materials uses a maximum voltage of 2.2 V, as opposed to the more conventionally used value of 2.45 V. It is possible that this choice of voltage window was determined by an earlier onset of electrolyte oxidation, causing the formation of the SEI film. X-ray diffraction (XRD) and Raman spectroscopy data from these publications revealed that the materials possess a relatively low degree of graphitic character, which would not allow intercalation of ions, thus leading to a primarily capacitive energy storage mechanism. It is likely that the low crystallinity is also the main cause of the reduction of the voltage window. This hypothesis is supported by the fact that amorphous or semi-graphitic carbons generally have a high amount of surface defects such as oxygen-containing functional groups (carbonyl, hydroxyl, carboxyl), non sp$^2$-hybridised carbon sites, and graphitic edges.\cite{chen_defect-free_2017} These defects could possibly act as “catalytic sites”, promoting the electrolyte degradation reaction. \cite{su_nanocarbons_2013,mao_nanocarbon-based_2014} In addition, the high surface area of the material could further exacerbate this process by drastically increasing the density of reaction centers in the material. Conversely, although the graphitic foam reported by Lin et al.\cite{lin_ultrafast_2015} is also characterized by a high surface area, XRD and Raman data indicate that the material possesses a highly ordered graphitic structure, with very few defect sites.\cite{chen_three-dimensional_2011} Thanks to these properties, the material allows an ultrafast reversible ion intercalation mechanism, without any significant electrolyte decomposition at potentials up to 2.45 V. This report is therefore also consistent with the hypothesis that surface defects are responsible for the early onset of electrolyte oxidation.

In order to test these hypotheses, a new carbonaceous material was employed as cathode for AIBs. Electrospun carbon nanofibers (CNF) are a versatile carbon material, featuring moderately high surface areas, and a tunable degree of graphitic character.\cite{inagaki_carbon_2012} They can be synthesized easily via electrospinning and carbonization of an appropriate precursor polymer; the most commonly used starting material for this purpose is polyacrylonitrile (PAN), thanks to its high carbon yield and convenient mechanical properties.\cite{nataraj_polyacrylonitrile-based_2012} Furthermore, similarly to pyrolytic graphite paper and graphitic foam, they can be used directly as self-standing electrodes, without the use of any binders or current collectors.

Two types of CNF sheets were fabricated, with different degrees of graphitic character. This was achieved by adding cobalt acetate tetrahydrate to one of the precursor solutions, which provided a source of metal catalyst, promoting the growth of graphitic carbon.\cite{krivoruchko_new_1998} The metal catalyst was then removed by washing the nanofiber mats with aqueous hydrochloric acid. A Brunauer-Emmett-Teller (BET) surface area measurement was also performed
on the non-doped fibers, yielding a moderately high surface area of about 47 m$^2$ g$^{-1}$.

\begin{figure}
  \centering
 \includegraphics[width=0.4\textwidth]{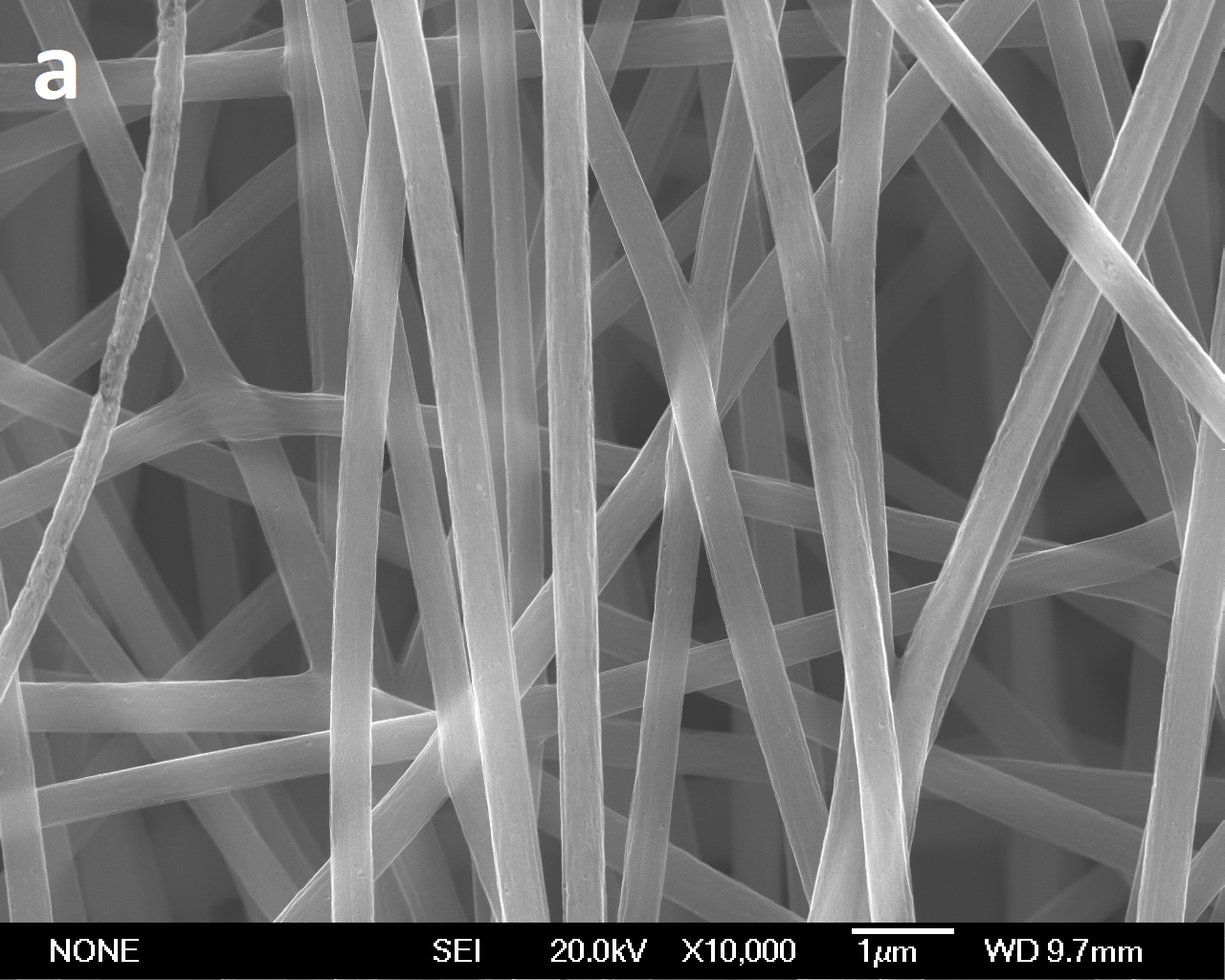}
 \includegraphics[width=0.4\textwidth]{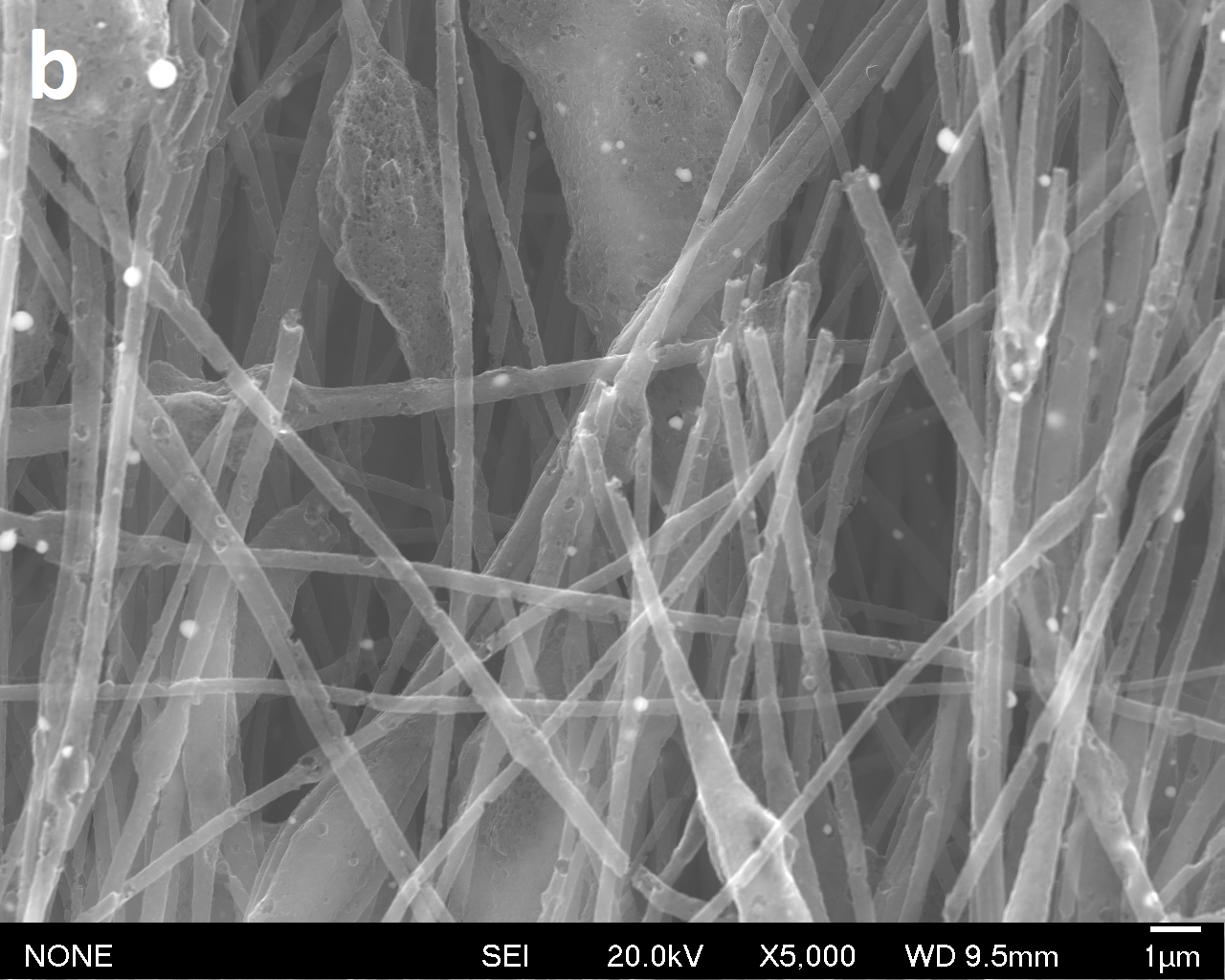}
  \includegraphics[width=0.4\textwidth]{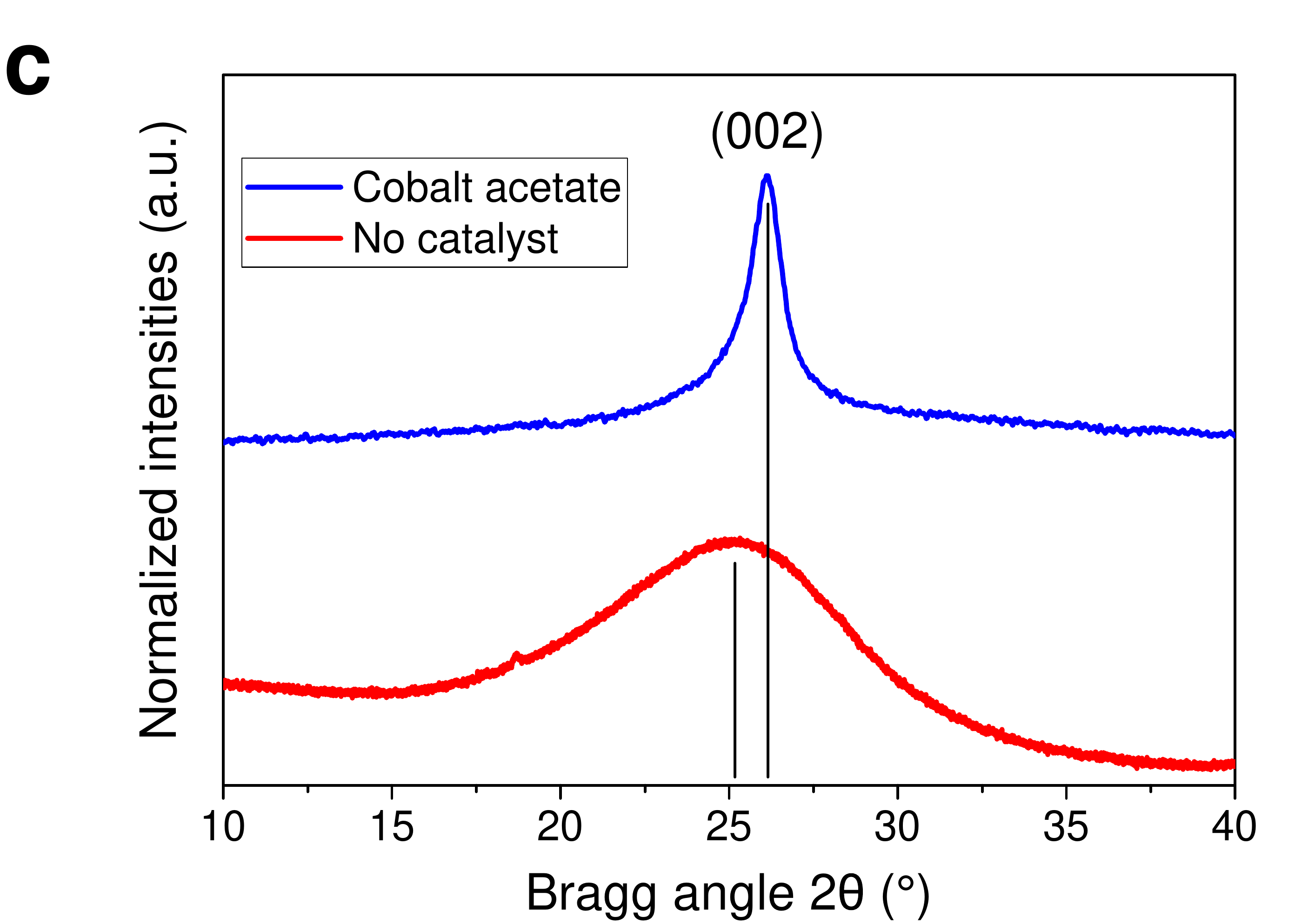}
   \includegraphics[width=0.4\textwidth]{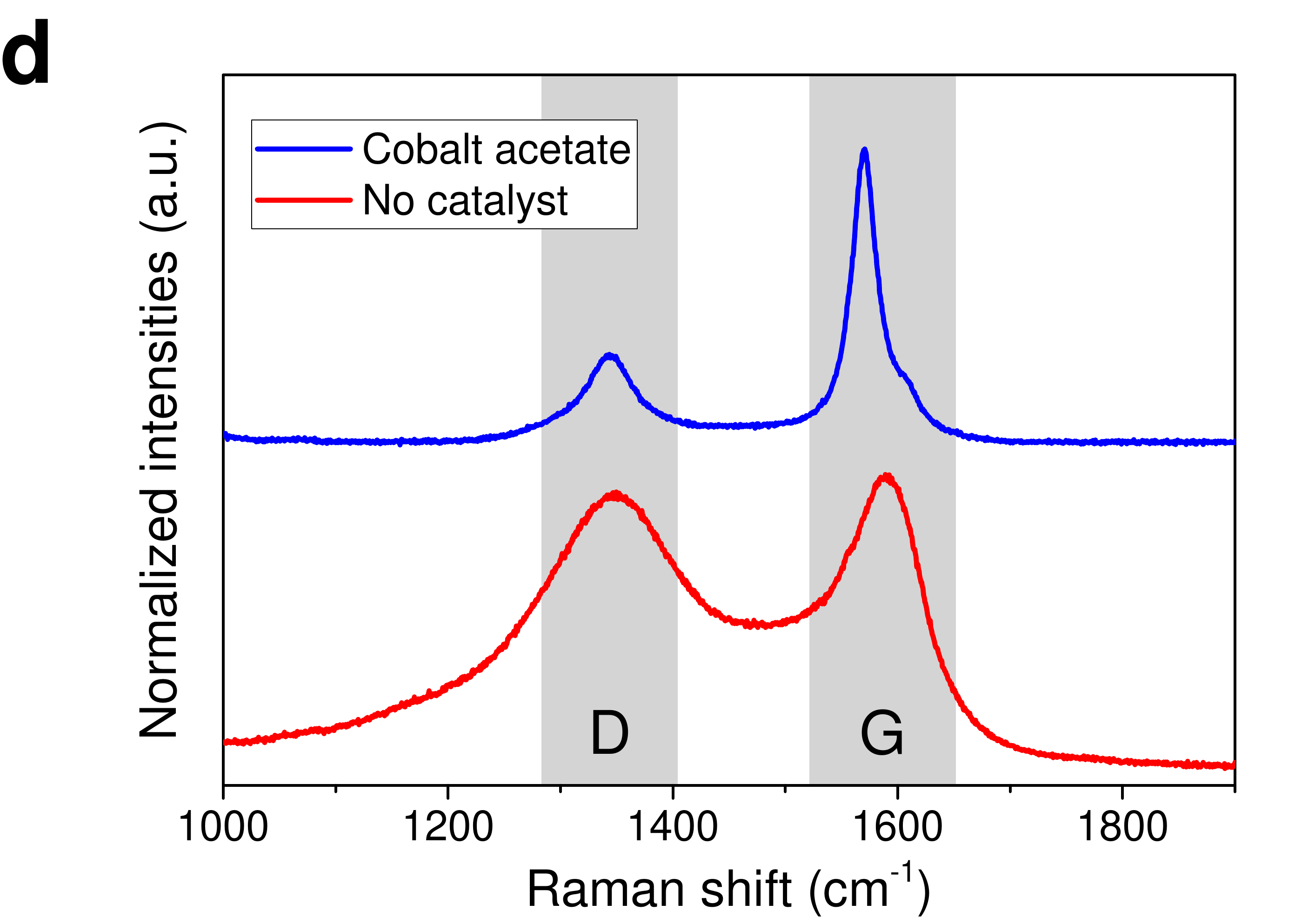}
  \caption{ Structural characterization of the two CNF samples: (a,b) SEM micrographs of the sample prepared without any catalyst (a) and with the addition of cobalt acetate (b); (c) X-ray diffractograms and (d) Raman spectra of the two CNF samples prepared.}
  \label{fig:fig2}
\end{figure}

Scanning electron microscopy (SEM) micrographs of the two CNF samples were acquired (Figure \ref{fig:fig2}). The images showed nanofibers of about 300-400 nm diameters in both cases, with the presence of a few bead-shaped irregularities, spherical cavities, and residual metal particles in the case of the cobalt acetate-doped fibers. It can be seen from the X-ray diffractogram in Figure \ref{fig:fig2}c that, with the addition of the metal catalyst, the characteristic (002) peak of graphite becomes notably narrower, and shifts from 25 to 26.1 degrees. This corresponds to an increase in the size of crystalline domains in the material, and a decrease in the interlayer spacing between the basal planes of graphite, in accordance with Bragg's law. The Raman spectra of the two samples in Figure 2d also revealed that the addition of metal catalyst caused a reduction in the overall width of the D and G bands, located at about 1350 and 1580 cm$^{-1}$ respectively. This indicates an overall increase in the crystallinity of the carbon material. Furthermore, the ratio between these peaks decreased, suggesting a reduction of the amount of defects in the material such as non-sp$^2$ hybridized carbon atoms, graphitic edges, dislocations, and elemental impurities.\cite{ferrari_raman_2007,malard_raman_2009} This data therefore indicates that the two samples were characterized by a different degree of crystallinity, with the cobalt-doped CNF achieving a notable increase in graphitic character.

\begin{figure}
  \centering
 \includegraphics[width=0.3\textwidth]{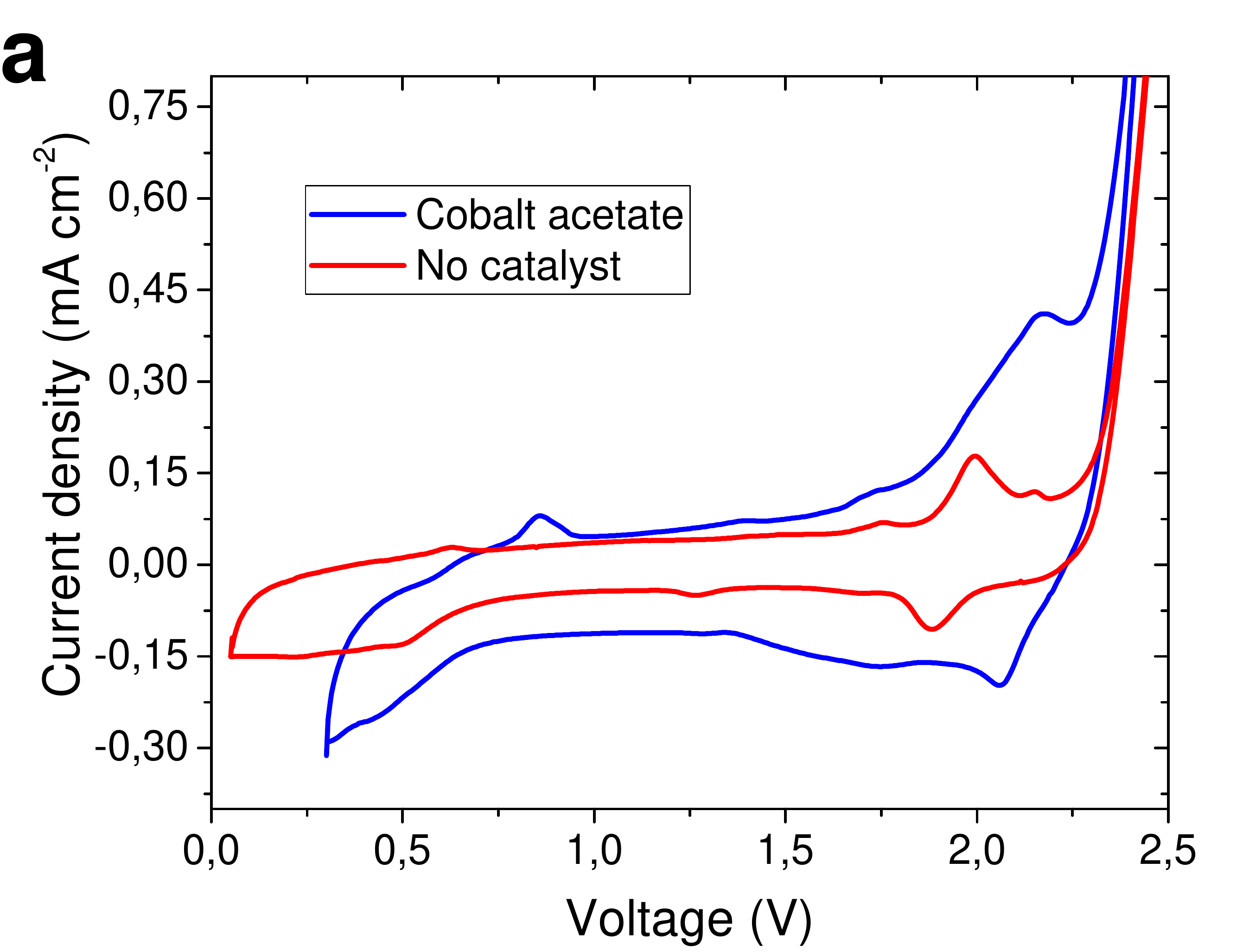}
 \includegraphics[width=0.3\textwidth]{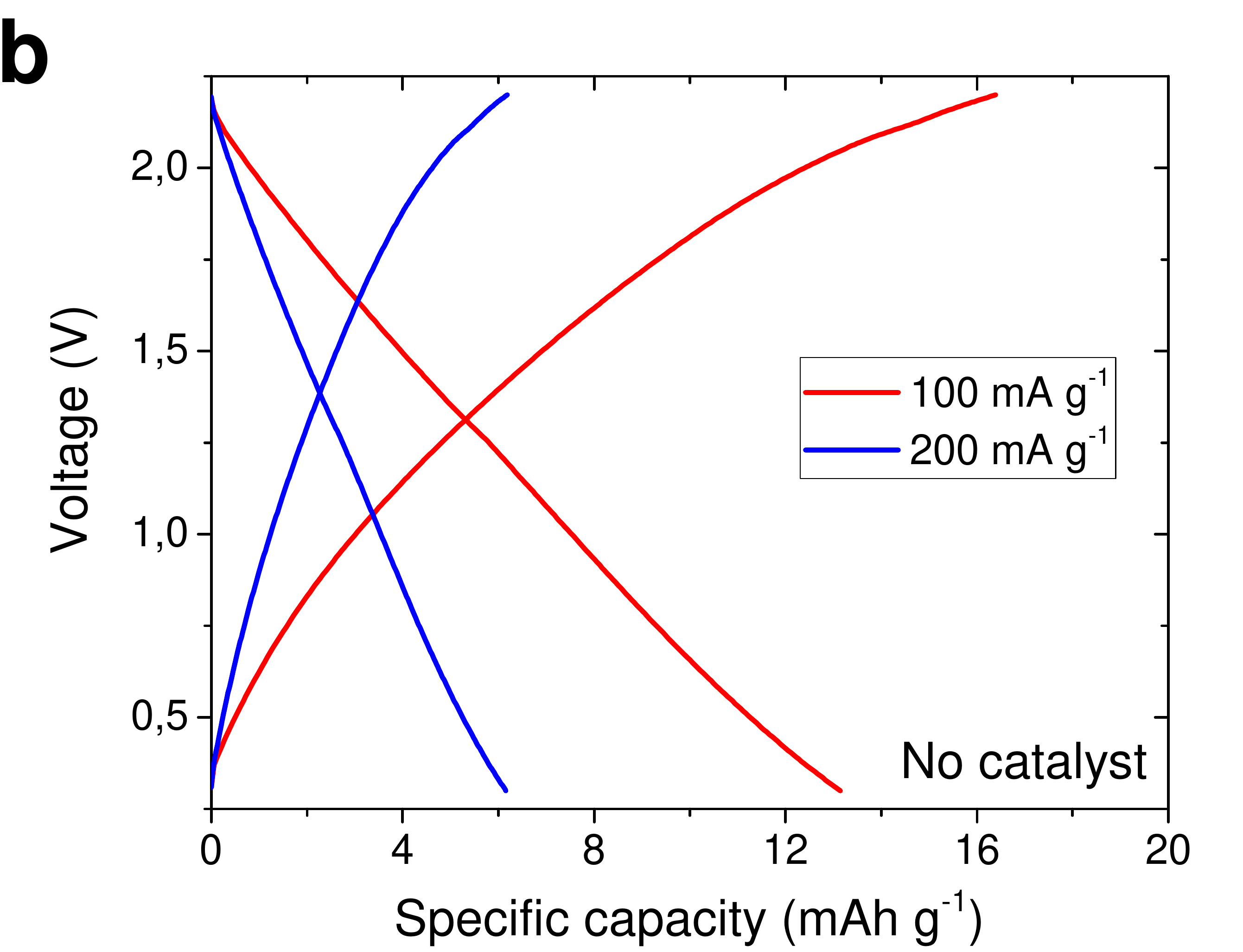}
  \includegraphics[width=0.3\textwidth]{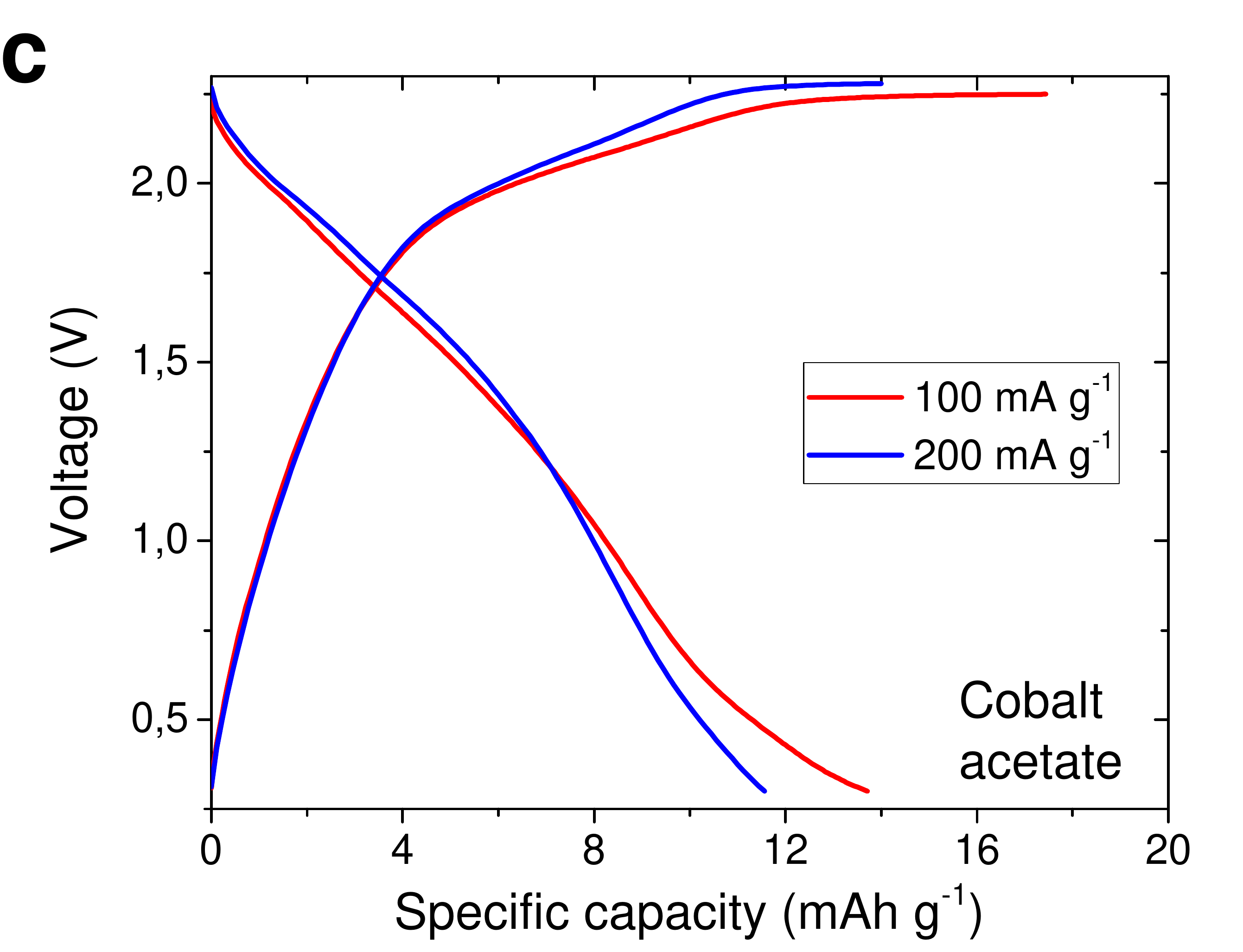}
   
  \caption{ Electrochemical performance of Swagelok-type cells built using the two CNF samples as cathodes: (a) Cyclic voltammograms (third cycle, 10 mV $^{-1}$), and (b) typical galvanostatic charge-discharge profiles (tenth cycle).}
  \label{fig:fig3}
\end{figure}

Swagelok-type cells were assembled using the self-standing CNF mats as cathodes for electrochemical testing. The cyclic voltammograms (CV) in Figure \ref{fig:fig3}a show the presence of a sharp increase in current above 2.2 V. This corresponds very likely to the bulk oxidation of the electrolyte, and appears to take place at a considerably lower potential than in PGP (See also Figure S2
). This also implies that, for galvanostatic charge-discharge tests, the conventional upper potential limit of 2.45 V is not applicable for this material, and a lower value must instead be used. 

A series of oxidation and reduction peaks can also be observed between about 1.7 and 2.2 V. These are in a similar range that is associated with the intercalation and deintercalation of AlCl$_4^-$ ions observed in graphitic materials. The peaks are more prominent in the plot relative to the more graphitic CNF, which is compatible with the prediction that a more crystalline material would allow a more effective intercalation mechanism, as suggested by other reports.\cite{kravchyk_efficient_2017,wang_kish_2017} However, it is also worth noting that both CVs feature a prominent hysteretic baseline, indicating the presence of a capacitive or pseudo-capacitive process.\cite{conway_transition_1991,arico_nanostructured_2005} It is therefore likely that, the primary energy storage mechanism in the device is in fact not the intercalation of AlCl$_4^-$ ions into the cathode, but the adsorption of the ions onto the electrode surface. This result indicates that only highly graphitic carbons enable the reversible intercalation of ions, and even a relatively low amount of structural disorder, as in the case of cobalt-doped CNFs, is enough to hinder the intercalation reaction considerably. The high surface area, on the other hand, could favor the adsorption of a large quantity of ions on the electrode surface, thus making the capacitive adsorption the predominant process in these devices. 

\begin{figure}
  \centering
 \includegraphics[width=0.4\textwidth]{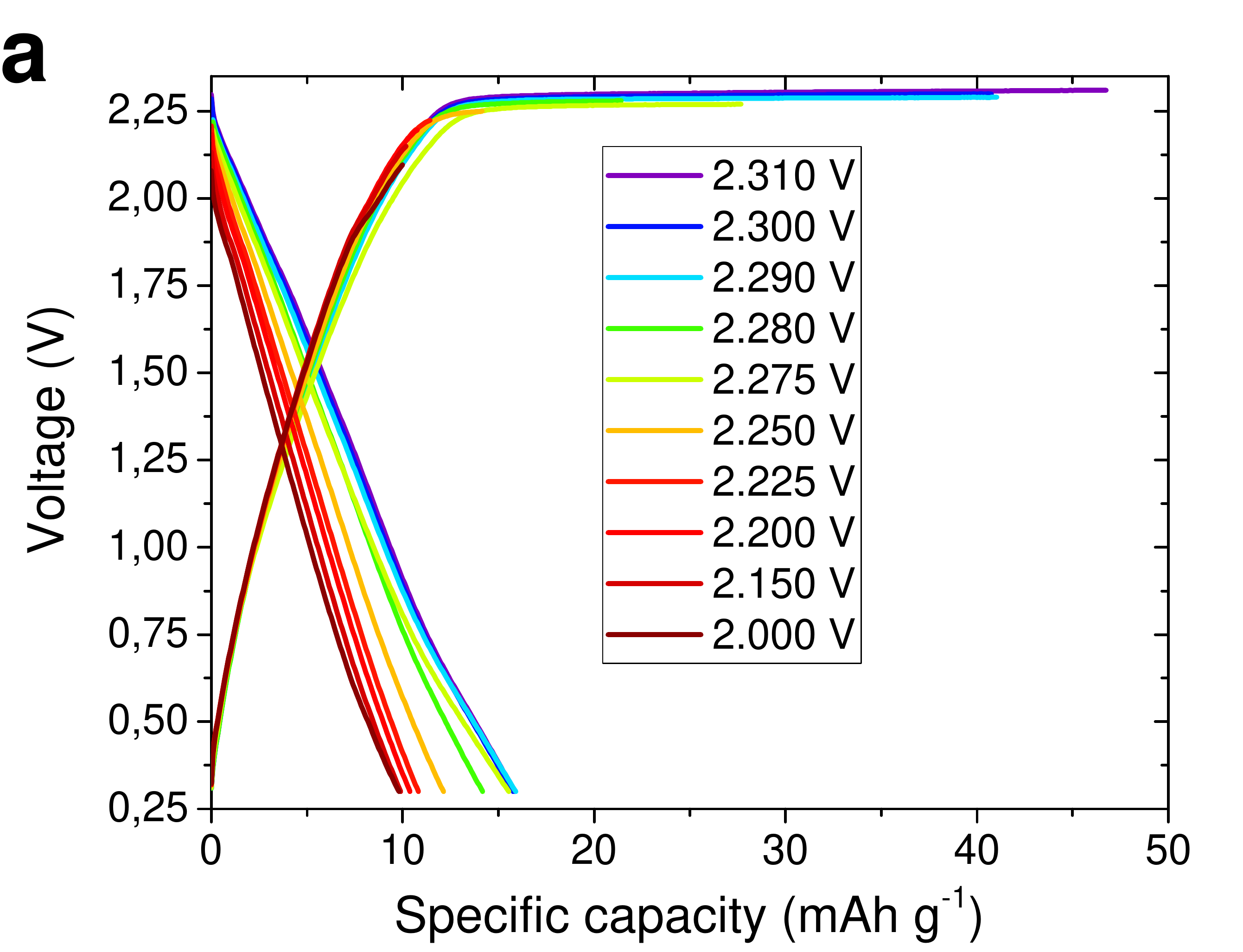}
 \includegraphics[width=0.4\textwidth]{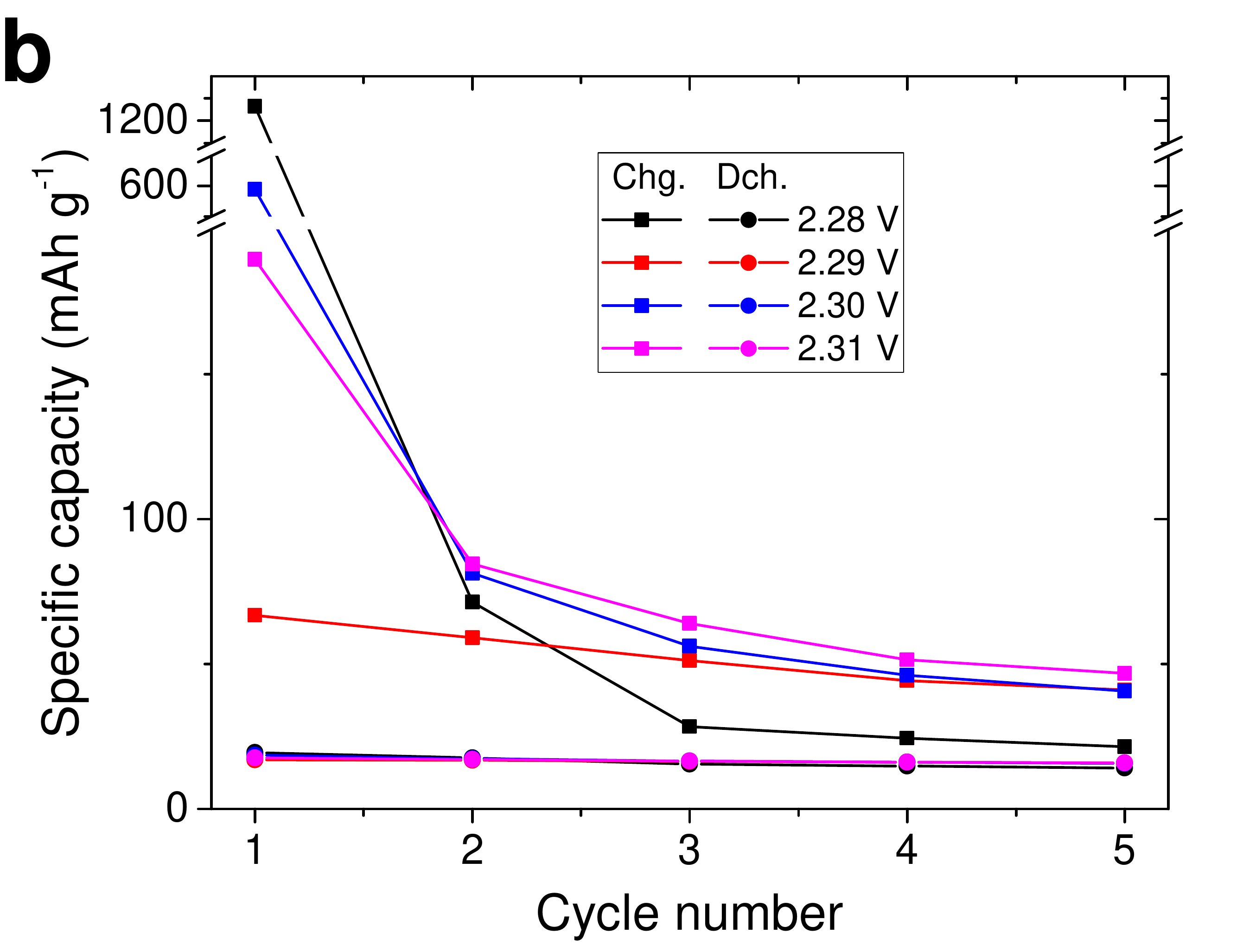}

  \caption{ (a) Galvanostatic charge-discharge profiles (fifth cycle, 100 mA g$^{-1}$) of a Swagelok-type cell built using CNF (non-catalyzed) as cathode, using increasing upper voltage limits. (b) Specific capacities for the first five cycles relative to the same galvanostatic charge-discharge test. }
  \label{fig:fig4}
\end{figure}

Galvanostatic cycling tests (Figure \ref{fig:fig3}b,c) also confirm the results observed in the CV experiments: the devices built using cobalt-doped CNFs yielded a marginally higher discharge capacity, and the corresponding galvanostatic profiles featured weakly defined plateaus at about 2.1 and 1.7 V for the charge and discharge curves, respectively, possibly indicating the presence of a more faradaic energy storage process. The poor specific capacities obtained for both materials, however, confirm that little to no ion intercalation takes place in these materials. It is also worth noting that the upper voltage limit had to be lowered considerably from the conventional value of 2.45 V to obtain a charging-discharging protocol with a somewhat reasonable efficiency. If a voltage cut-off higher than 2.25 V was used, the specific capacity of the charging step increased dramatically, but a negligible increase of discharge capacity was observed (Figure \ref{fig:fig4}a). One possible explanation for this behavior is that a small quantity of ion intercalation could indeed happen in CNFs, but the voltage overlap between this reaction and the degradation of the electrolyte was such that the desired faradaic process could not take place. Another interesting observation is that when higher voltage cut-offs were used, charging capacities tended to progressively decrease with every cycle, resulting in an increase in coulombic efficiency (Figure \ref{fig:fig4}b). This is another point of similarity with the behavior of carbonaceous materials as anodes for lithium-ion batteries, and is consistent with the assumption that surface defects are responsible for the decomposition of the electrolyte at lower voltages: as the SEI film is formed and increases in thickness, the surface of the cathode gets progressively shielded from direct contact with the electrolyte, making the decomposition reaction gradually less prominent.

\begin{figure}
  \centering
 \includegraphics[width=0.3\textwidth]{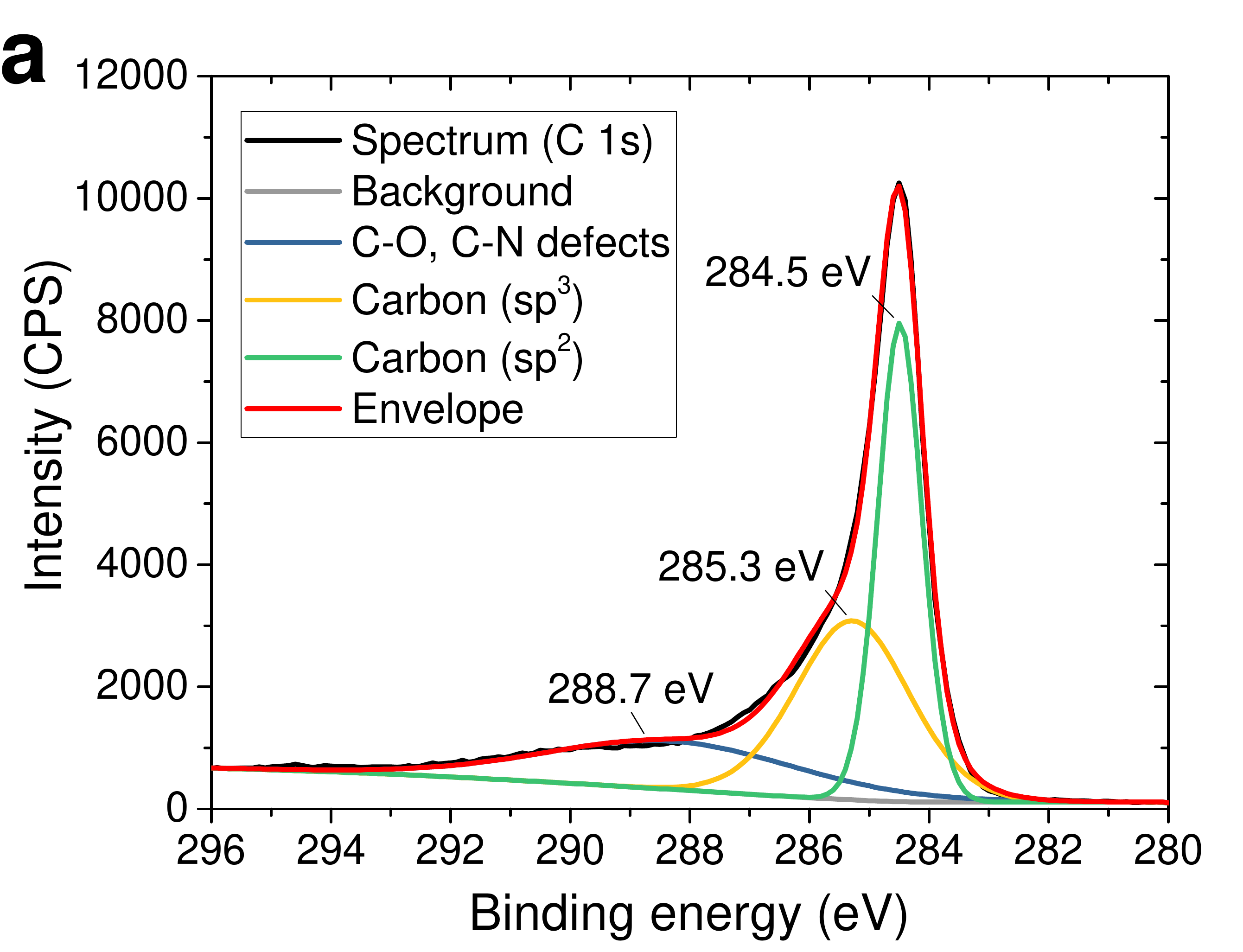}
 \includegraphics[width=0.3\textwidth]{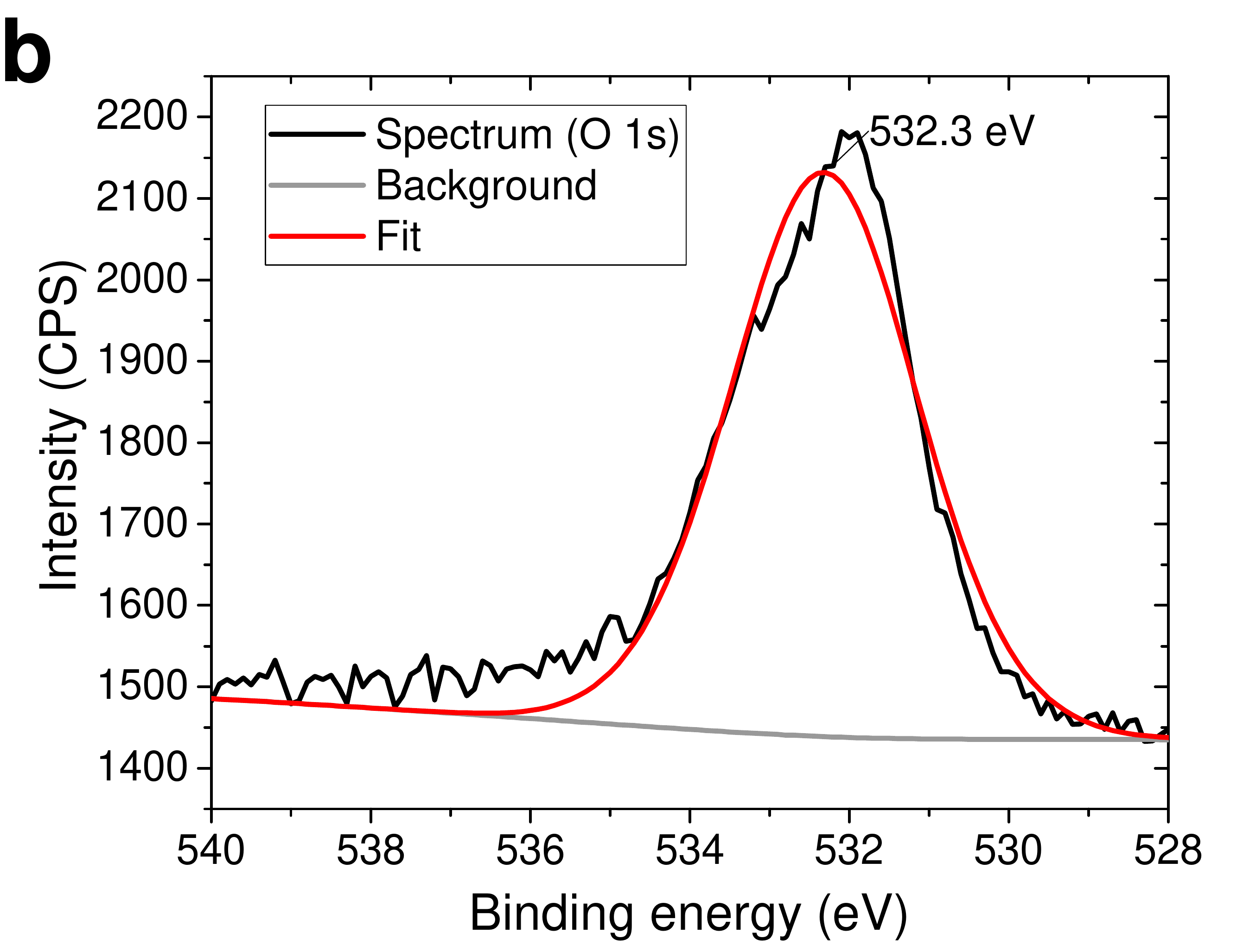}
\includegraphics[width=0.3\textwidth]{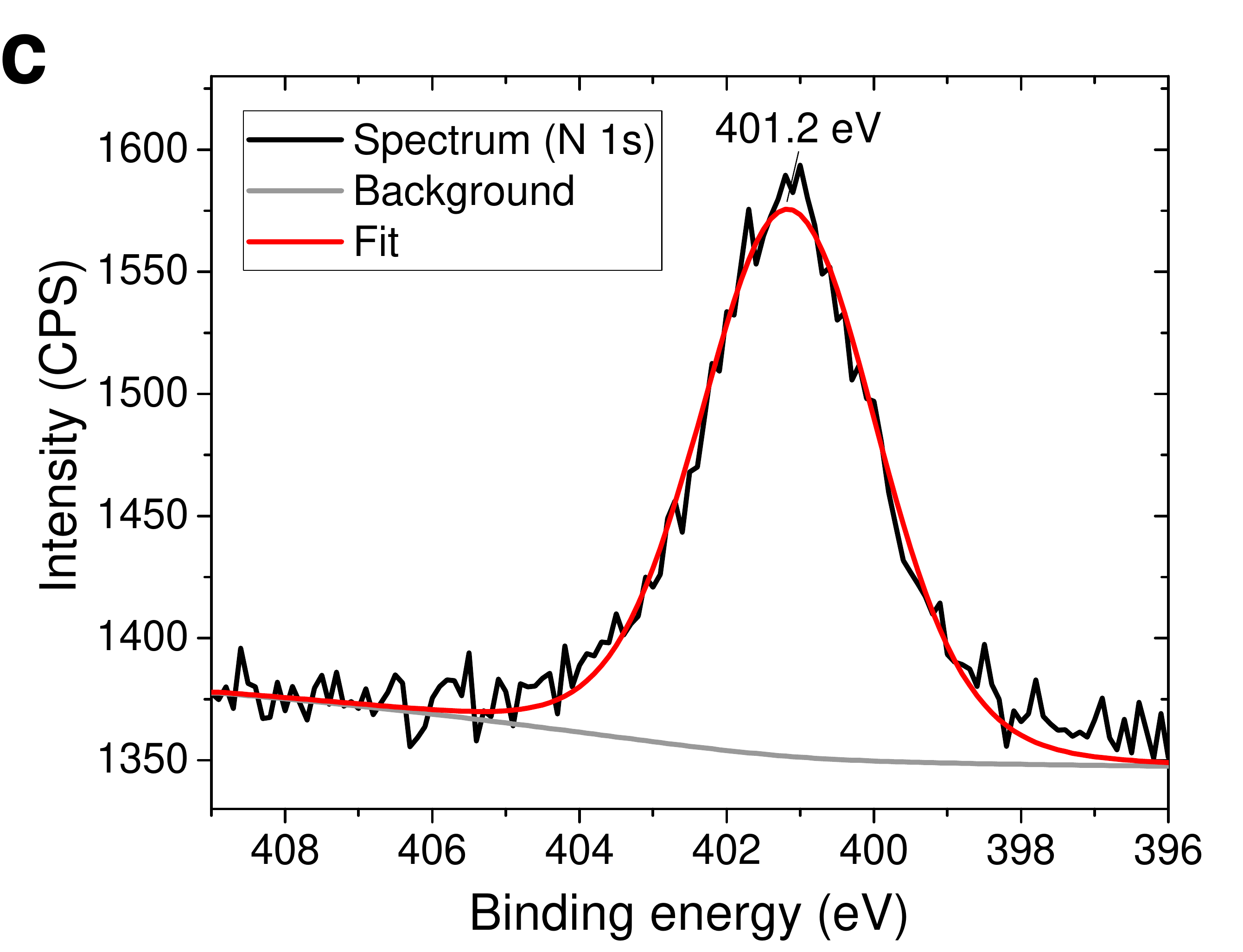}

  \caption{ Deconvoluted XPS narrow scans of the most prominent spectral regions for a pristine CNF (non-catalyzed) sample: C 1s (a), O 1s (b), and N 1s (c). }
  \label{fig:fig5}
\end{figure}

X-ray photoelectron spectroscopy (XPS) was used to study the surface composition of CNFs and quantify the presence of defects in the materials. It can be seen from Figure \ref{fig:fig5}a that the C 1s narrow scan shows a peak with highly asymmetrical line shape, indicating the presence of multiple chemical environments. The spectrum was deconvoluted using a multiple peak fit, revealing the presence of three main components: the most intense peak, with a maximum at 284.5 eV, can be associated with bulk sp$^2$ carbon, likely the main component of the nanofibers. The second peak, with a maximum at 285.3 eV, is imputable to the presence of irregular or defective sp$^3$ carbon, which is consistent with our observations from XRD and Raman experiments. The position of these peaks is also in good agreement with other reports of similar carbon materials.\cite{merel_direct_1998,filik_xps_2003} Finally, a less intense and notably broader peak, centered at 288.7 eV, can also be found. This peak is likely ascribable to the presence of carbon atoms bound to different, more electronegative elements such as carbon and nitrogen. Previous literature indicates that this component is associated with the presence of functional groups such as hydroxyl, ether, nitrile, carbonyl, carboxyl or ester.\cite{moulder_handbook_1993,zhou_characterization_2007} The presence of heteroatomic defects is also supported by the narrow scans for the O 1s and N 1s regions in Figure \ref{fig:fig5}b and Figure \ref{fig:fig5}c, respectively: two fairly symmetrical peaks, in the binding energy range corresponding to the reported values for the aforementioned functional groups, can be found in these two regions.\cite{moulder_handbook_1993,zhou_characterization_2007,beamson_high_1992} XPS data therefore provides additional evidence for the presence of heteroatomic defects in the materials, which could potentially be responsible for unwanted side reactions with the chloroaluminate species in the electrolyte, causing the formation of the SEI film.

\begin{figure}
  \centering
 \includegraphics[width=0.7\textwidth]{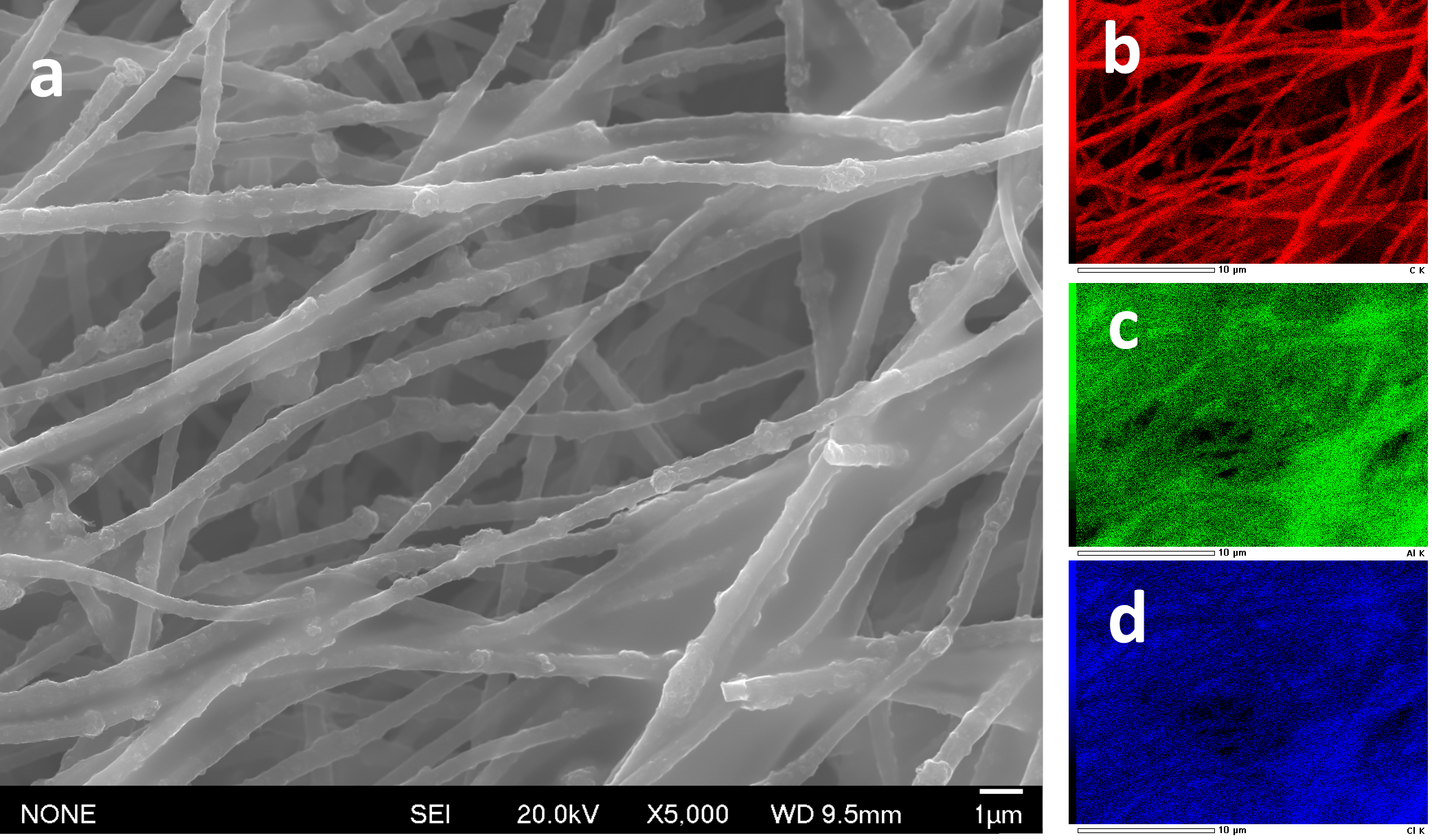}

  \caption{ SEM micrograph (a) and EDXS elemental maps of carbon (b), aluminum (c), and chlorine (d) of CNF (non-catalyzed) after a galvanostatic charging step (100 mA g$^{-1}$ for approximately 10 hours).}
  \label{fig:fig6}
\end{figure}

\begin{figure}
  \centering
 \includegraphics[width=0.7\textwidth]{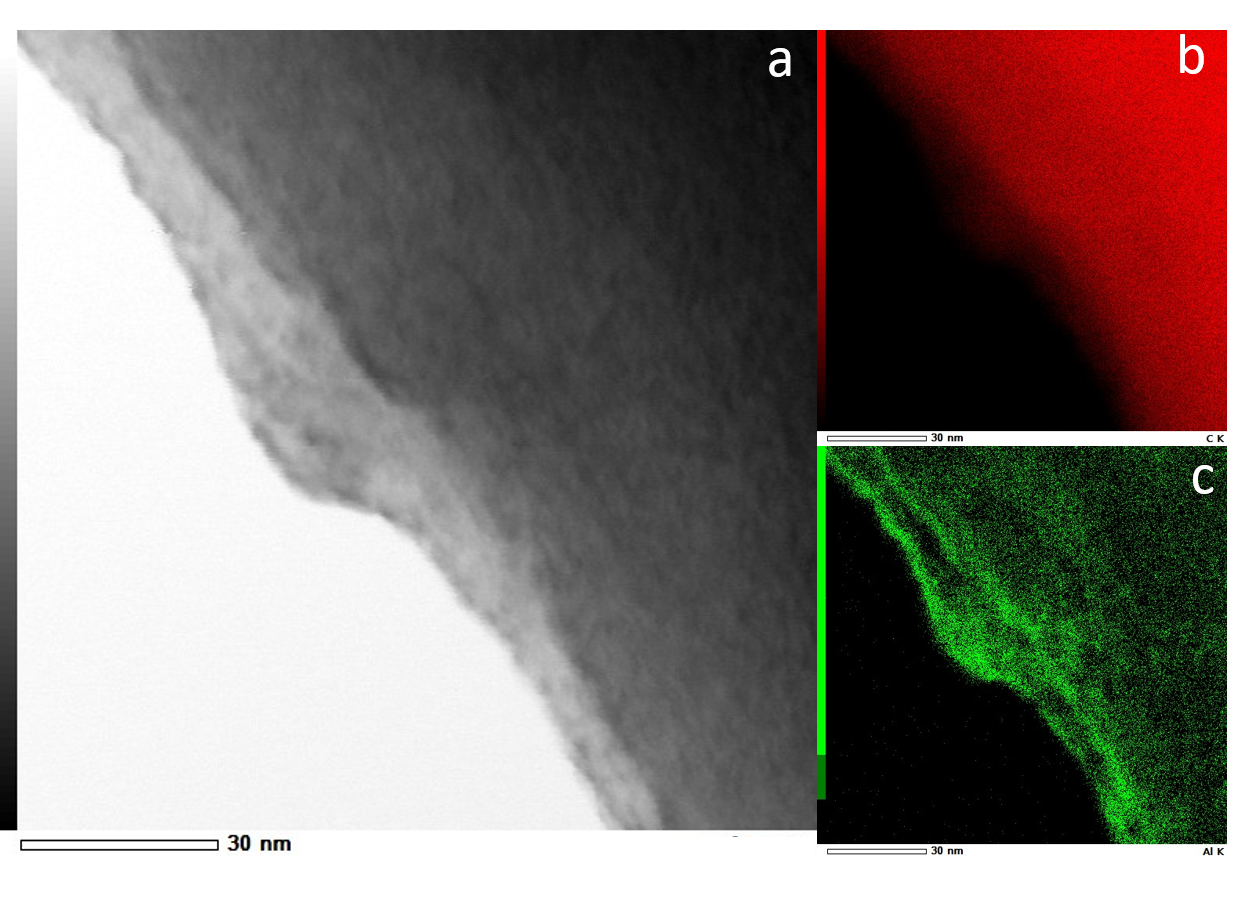}

  \caption{ (a) STEM micrograph of an individual CNF (non-catalyzed) after a galvanostatic charging step (100 mA g$^{-1}$ for approximately 10 hours). A layer approximately 5nm thick is observed. (b) Carbon EDXS elemental map, showing that little to no carbon is present in the SEI layer. (c) Aluminium EDXS elemental map.}
  \label{fig:fig7}
\end{figure}

In order to investigate the formation process and the composition of the SEI, a cell using non-catalyzed CNFs as cathode was charged using a galvanostatic current of 100 mA g$^{-1}$ for approximately 10 hours, without the use of voltage cut-offs. The battery was then disassembled, and the cathode was analyzed using SEM and EDXS elemental mapping. It can be seen from Figure 6 that the same kind of lumpy film observed for PGP (Figure S1
, containing aluminum and chlorine species, was found on the surface of the nanofibers after the electrochemical process. Transmission electron microscopy was also used to investigate the film further (Figure \ref{fig:fig7}). A layer of approximately 5nm is observed. This layer contains no carbon but is aluminum and chlorine rich suggesting that electrolyte composition has taken place.

It is therefore evident that the SEI film was indeed formed as a consequence of electrochemical processes taking place at the cathode during the initial charging step. These processes are likely related to the oxidizing potential applied to the material during the galvanostatic process, which could cause the decomposition of some of the species in the electrolyte.

\begin{figure}
  \centering
 \includegraphics[width=0.4\textwidth]{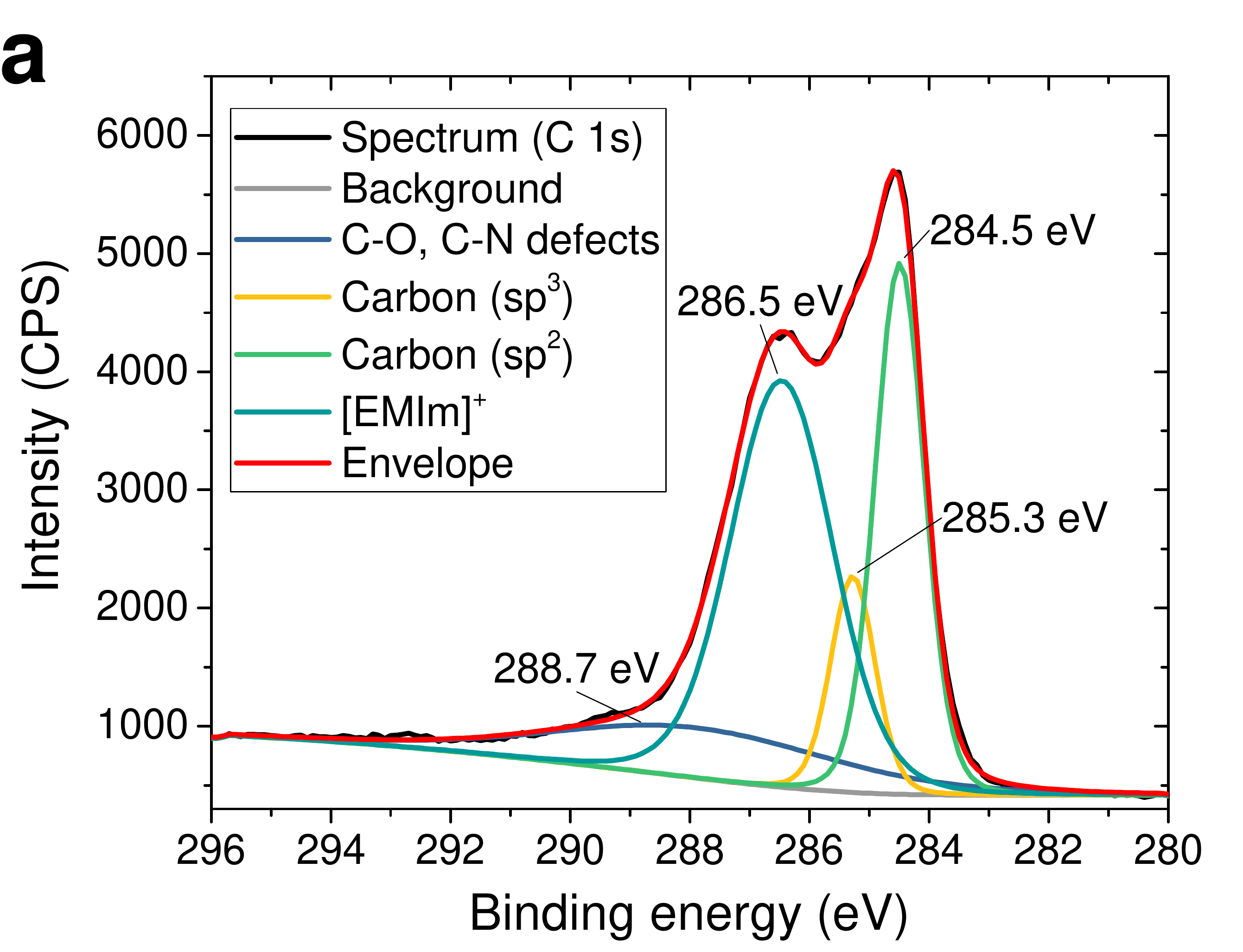}
  \includegraphics[width=0.4\textwidth]{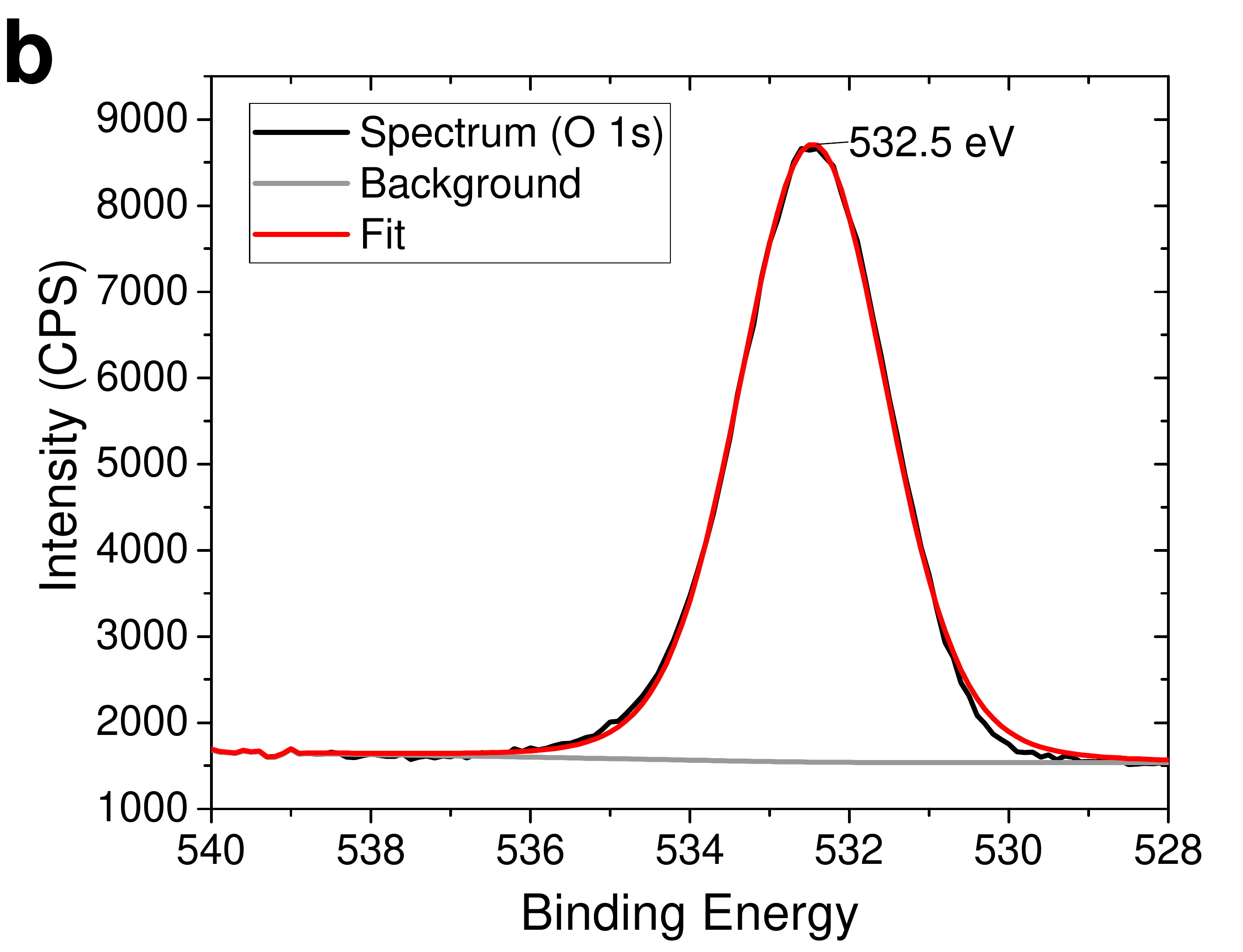}
   \includegraphics[width=0.4\textwidth]{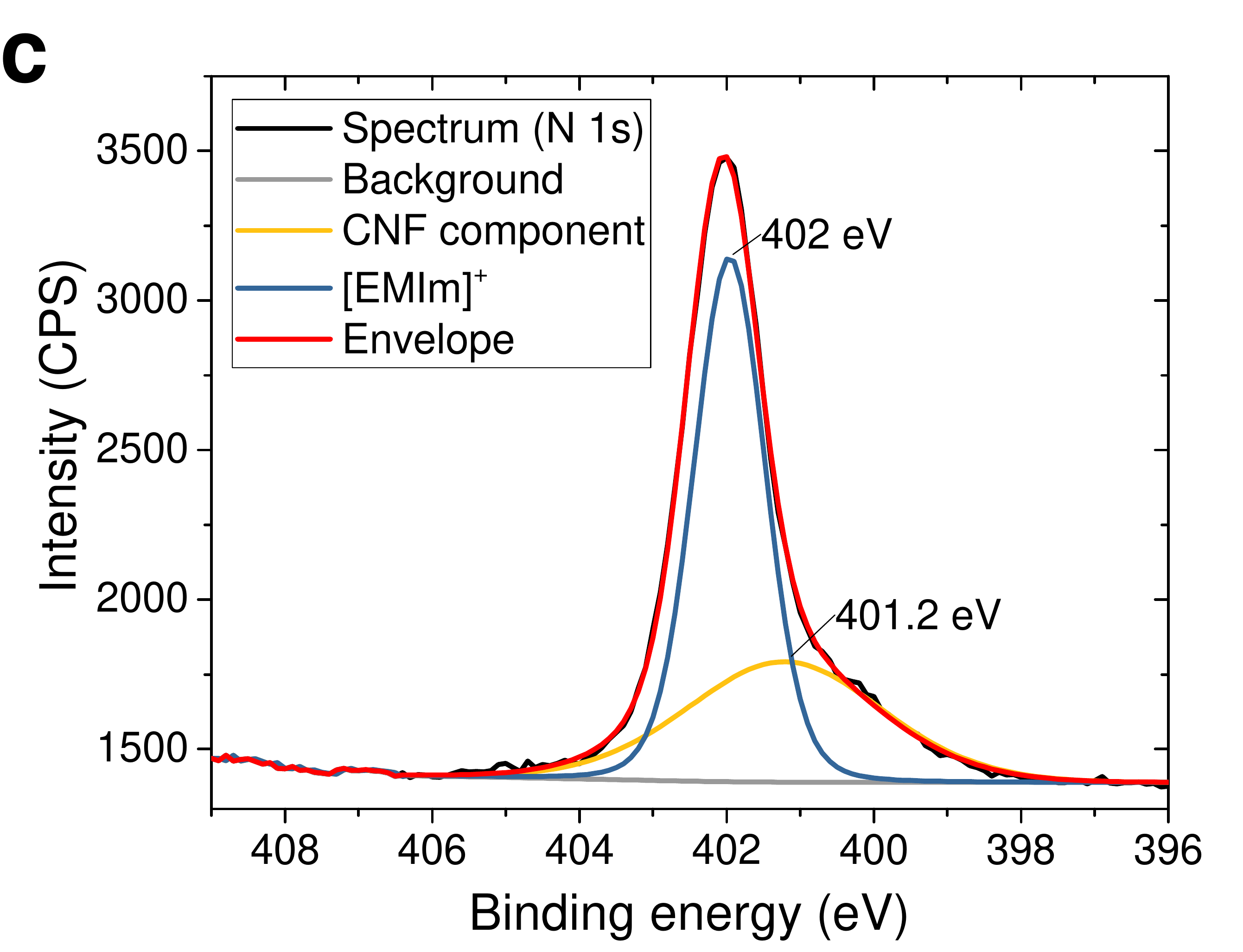}
    \includegraphics[width=0.4\textwidth]{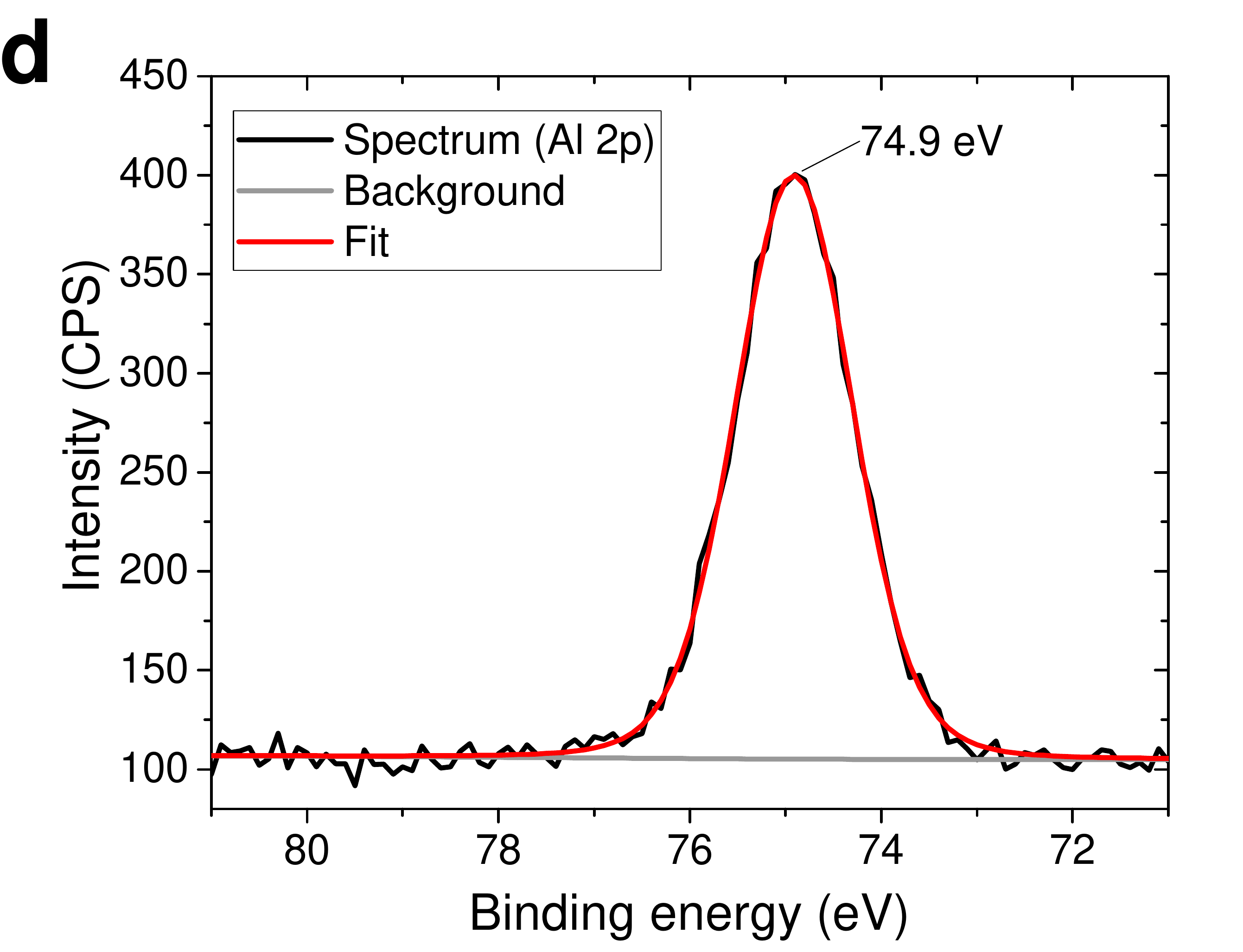}
     \includegraphics[width=0.4\textwidth]{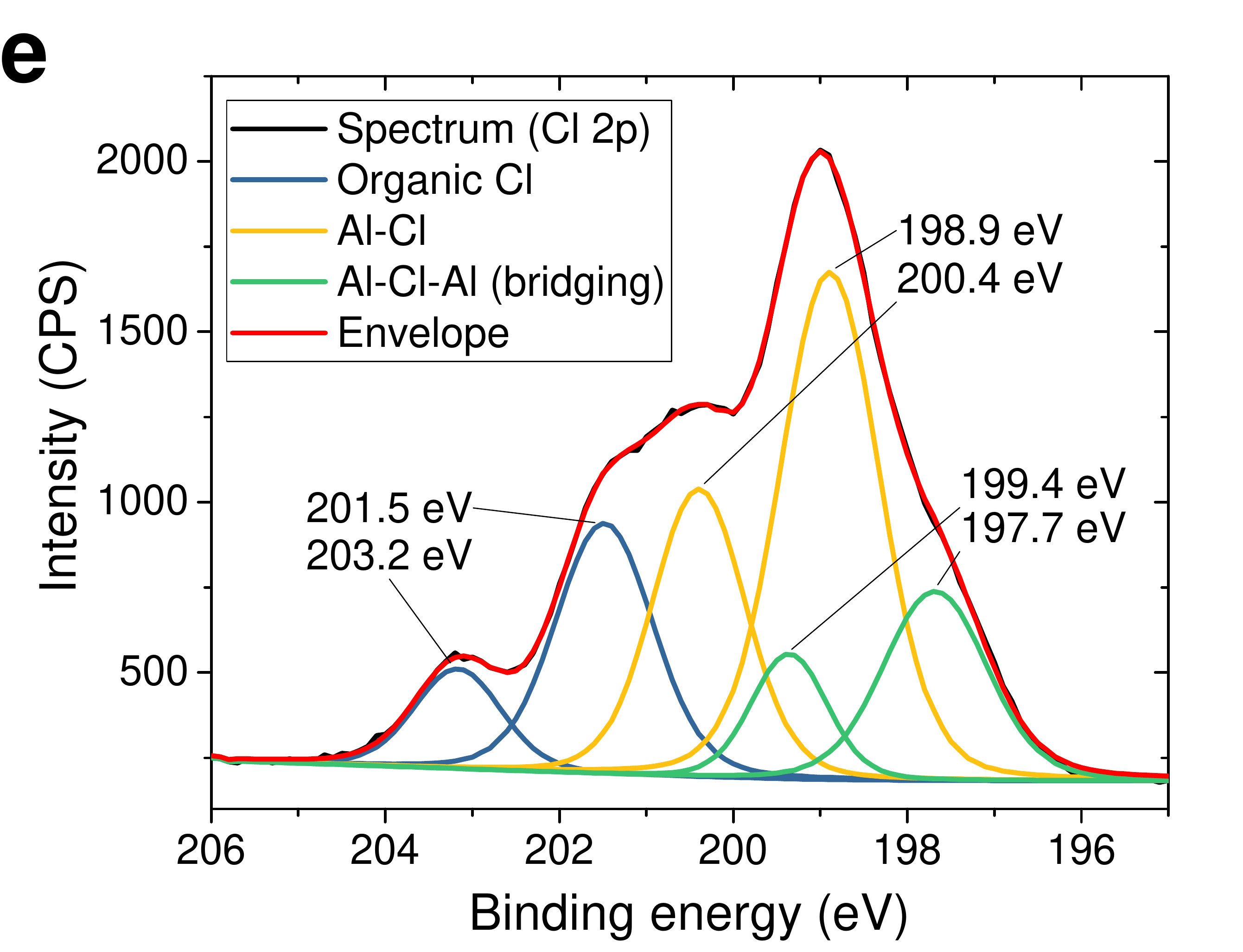}

  \caption{ Deconvoluted XPS narrow scans of selected spectral regions for a CNF sample (non-catalyzed), after a galvanostatic charging step (100 mA g$^{-1}$ for approximately 10 hours): C 1s (a), O 1s (b), N 1s (c), Al 2p (d), and Cl 2p (e).}
  \label{fig:fig8}
\end{figure}

XPS was also performed on the charged CNFs sample to further study the composition of the SEI and tentatively identify any new compounds created by electrolyte decomposition. The C 1s narrow scan in Figure \ref{fig:fig8}a shows that, in addition to the same type of carbon environments observed for pristine CNFs (as indicated by the presence of peaks matching the positions of the peaks found in the pristine CNF samples, with comparable linewidth), a new prominent peak, with a maximum at 286.5 eV, was found in the deconvoluted spectrum. This peak matches well with the reported binding energies for the carbon atoms in alkyl-imidazolium cations,\cite{caporali_x-ray_2006,lockett_angle-resolved_2008} suggesting that either the [EMIm]$^+$ ion or some of its derivatives are likely present in the film. This hypothesis is further supported by the N 1s narrow scan in Figure \ref{fig:fig8}c, as a new peak centered at 402 eV was found. This value is also in good agreement with literature reports of N 1s photoelectron spectra for alkyl-imidazolium compounds.\cite{caporali_x-ray_2006,weingarth_electrochemical_2012} Interestingly, similar photoelectron spectra were acquired for charged graphite samples by a recent publication, but a different interpretation was proposed by the authors: the C 1s peak at 286.5 eV was attributed to either oxidized graphite (due to anion intercalation), or partial co-intercalation of [EMIm]$^+$ cation in the crystal lattice. However, given that the chemical information provided by XPS is highly surface-weighted, the hypothesis that such peaks are relative to species present on the surface of the sample (i.e. the SEI film) is more likely. Looking at the O 1s spectrum (Figure \ref{fig:fig8}b), no particular new features were found. The intensity of the main peak, however, was notably increased, suggesting the presence of new compounds, likely emerging from reactions involving atmospheric oxygen. Due to the large variety of plausible chemical species responsible for this peak, a clear interpretation of this phenomenon is not easy to produce. A possible explanation is the presence of decomposition products deriving from the chloroaluminate ions, such as aluminum oxides or hydroxides in the samples.\cite{powell_x-ray_1989,carlin_electrochemistry_1996} This hypothesis is also  supported by the presence of a peak in the Al 2p spectrum, centered around 74.9 eV (Figure \ref{fig:fig8}d). Possible compounds matching this range of binding energies include AlCl$_3$, aluminum oxides, and aluminum hydroxides.\cite{powell_x-ray_1989,carlin_electrochemistry_1996} The most important piece of information can be deduced from the Cl 2p spectrum in Figure \ref{fig:fig8}e: the highly irregular shape of the peak suggests the presence of multiple chemical states for chlorine. A curve fitting study revealed that the peak can be deconvoluted into three different chemical environments, as shown by the presence of three pairs of spin-orbit split peaks, featuring area ratios of approximately 1:2 and an energy splitting of about 1.6 eV.\cite{moulder_handbook_1993} While the two environments situated at lower energies could be attributed to the Cl atoms in either terminal or bridging positions of the chloroaluminate ions, as suggested by the binding energy values for other metal chlorides in the literature,\cite{moulder_handbook_1993,powell_x-ray_1989} the higher energy doublet falls in the typical range of organic (alkyl- or acyl-) chlorides. It is worth noting that none of the species initially present in the electrolyte contain a C-Cl bond. This indicates that an unforeseen side reaction in the electrochemical process must have caused the formation of such bonds. Therefore, one of the possible electrolyte degradation mechanisms likely involves the formation of carbon-chloride bonds between the chloroaluminate ions and either CNFs or the [EMIm]$^+$ cation.

Based on this data, it is therefore likely that both chloroaluminate species and their decomposition products are also present in the SEI films. A previous publication reports that, when high potentials (> 2.5 V vs Al$^{3+}$/Al) were applied to chloroaluminate electrolytes, the main mechanism of degradation is the oxidation of AlCl$_4^-$ ions generating chlorine gas, according to Equation \ref{eq:eq1}:\cite{jiang_electrodeposition_2006}

\begin{equation}
4 AlCl_4^- \rightarrow 2 Al_2 Cl_7^- + 2 e^- + Cl_2 \uparrow
\label{eq:eq1}    
\end{equation}

Assuming that this equation is correct, no new compounds, which could contribute to the formation of a solid film on the electrode, would be generated at high potentials. It is worth noting, however, that this equation refers to the oxidation of those species using a metallic tungsten electrode, with no organics or reactive functional groups present. On the other hand, numerous functional groups are present in the CNF electrodes used in our experiments. Furthermore, because the chloroaluminate species present in the electrolyte (especially Al$_2$Cl$_7^-$) are known to exhibit Lewis acid behavior,\cite{lungwitz_determination_2012,taylor_acidity_2013} they could form adducts with the functional groups of CNFs, serving as Lewis bases. Such adducts could further react by means of the oxidizing potential applied to the electrode, giving rise to new, insoluble compounds, contributing to the formation of the SEI film. Therefore, the oxide and hydroxide compounds detected in the XPS spectra could either be directly formed by reactions between the chloroaluminate ions and the surface functionalities of CNFs during the galvanostatic charging process, or as a further reaction of intermediate species present in the SEI film to ambient atmosphere. Similarly, the presence of XPS peaks ascribable to imidazolium ions or its derivatives is justified by two possible hypotheses: 1) residual [EMIm]$^+$ cations trapped in the SEI film, or 2) decomposition of these ions during the charging process. Such reactions could be caused by one or more of the aforementioned factors, i.e. the presence of Lewis acids/bases and relative adducts, oxidizing potentials applied to the cathode, and later exposure of the SEI film to atmospheric oxygen and water.

A control experiment was performed to verify the effect of the surface area to the electrochemical performance of CNFs. A monolithic, low-surface area (3.8 m$^2$ g$^{-1}$ according to BET measurements) carbon "film" (Figure S4
) was fabricated by drop-casting a PAN solution onto a glass surface, evaporating the solvent, and performing the same heat treatments used to prepare CNFs. Due to this fabrication process, the material is  composed of the same type of semi-graphitic carbon as CNFs; the surface of the material should also present the same type of defects, but its monolithic structure gives it a much lower surface area compared with the nanofibrous morphology of CNFs. After fabricating the carbon film, a cell was assembled using the material as cathode, and electrochemical testing was performed. It can be seen in the cyclic voltammogram in Figure S5
a that the current densities were generally lower compared with what was observed for CNFs (Figure \ref{fig:fig3}a). More importantly, the hysteretic baseline and increase in current after 2.2 V were also notably less pronounced. This result suggests that the reduced surface area of the film caused a lower capacitance in the material, but also an increase of the electrolyte decomposition potential. This result therefore confirms that the high surface area of CNFs was indeed a factor contributing to the narrowing of the useful potential window. This was further confirmed by the galvanostatic charge-discharge profiles in Figure S5
b: when cycling the device at 100 mA g$^{-1}$, the charging phase quickly reached the upper voltage limit of 2.45V. Even at a much lower charging rate of 10 mA g$^{-1}$, the charging step reached the target voltage after a reasonable charging capacity. On the other hand, no significant specific capacity was achieved in either case. This finding is consistent with the different structural properties of the material: the low surface area of the carbon film correlates to a lower amount of defect sites exposed to the electrolyte, making the decomposition process less prominent. In addition, the significant loss in specific capacity compared to CNFs also implies that capacitive surface ion adsorption is likely the predominant energy storage mechanism in both materials, as a reduction of surface area correlates with a reduction in the specific capacity in the cathode materials. 

The experiments described above suggest that surface defects were the main contributor to the reduction of the useful electrolyte window, causing the formation of the SEI. It is worth noting that even PGP, despite being highly graphitic (Figure S6
a), still likely possesses a certain amount of surface defects, as indicated by the presence of a weak D band in its Raman spectrum (Figure S6
b). Therefore, even in this case, the formation of an SEI can be justified by the presence of defects in the material. 

The theory that surface defects are responsible for the formation of an SEI was further tested with another control experiment: in order to reduce the accessibility of the electrolyte to the surface defects present in the cathode, a non-catalyzed CNF mat was dip-coated in a solution containing poly(vinylidene fluoride) (PVDF). Then, a battery device was assembled using the PVDF-coated CNF as the sample, and electrochemical testing was performed. This test was also conceptualized as a means to understanding the discrepancies in the reports of charge-discharge efficiencies in the first cycle between different reports: some publications report the use of carbonaceous cathodes without the use of any binder polymer\cite{kravchyk_efficient_2017,wang_kish_2017,antonioelia_insights_2017,stadie_zeolite-templated_2017}, while in other cases the material is cast as a film into a current collector, using binder polymers such as PVDF.\cite{wang_advanced_2017} The use of a binder polymer in those reports could have had an unforeseen effect on the electrochemical performance of the material: by effectively coating the carbon particles, it effectively shields the electrolyte from direct contact with the catalytic sites on their surface, thus inhibiting the undesirable decomposition reaction. 

\begin{figure}
  \centering
 \includegraphics[width=0.4\textwidth]{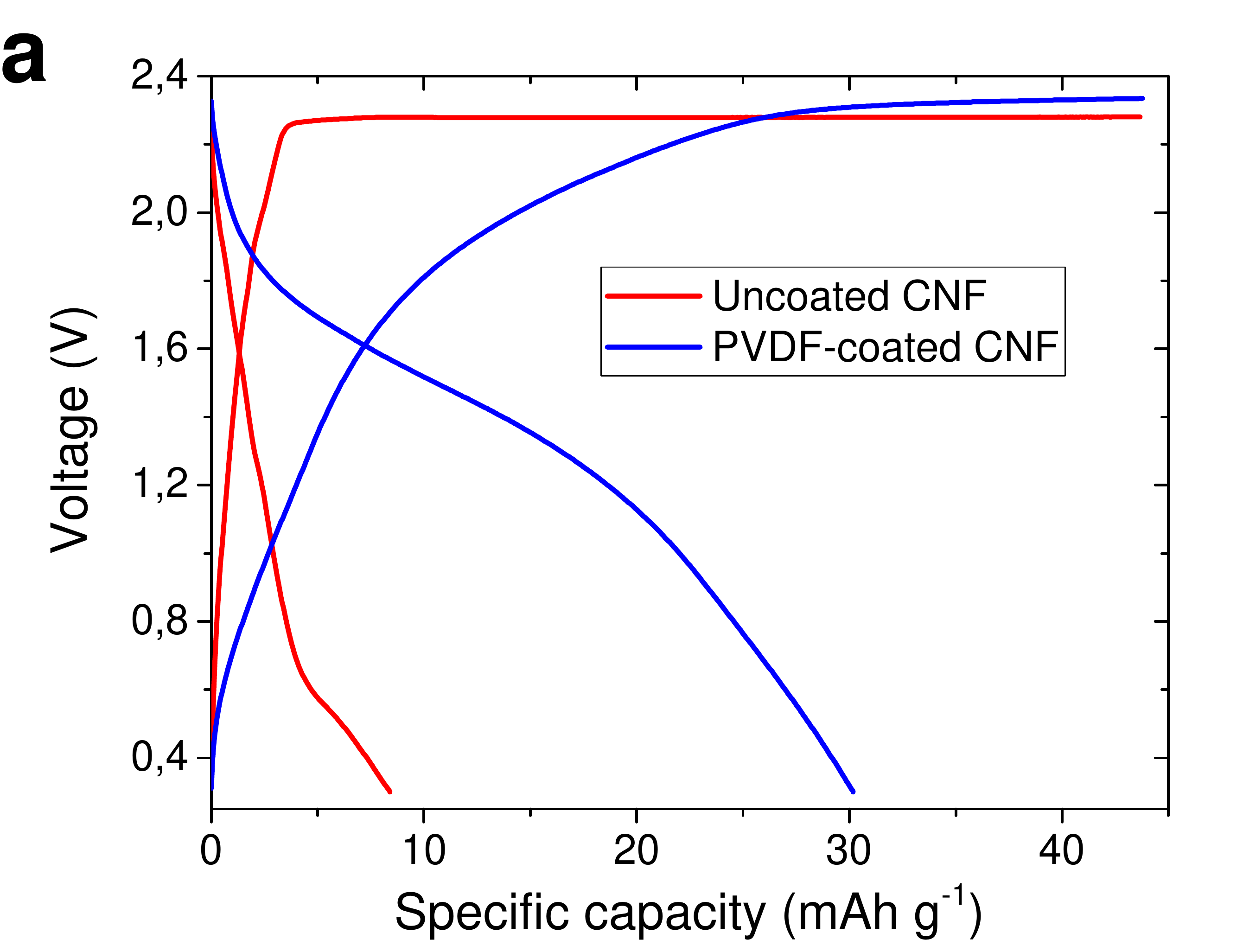}
  \includegraphics[width=0.4\textwidth]{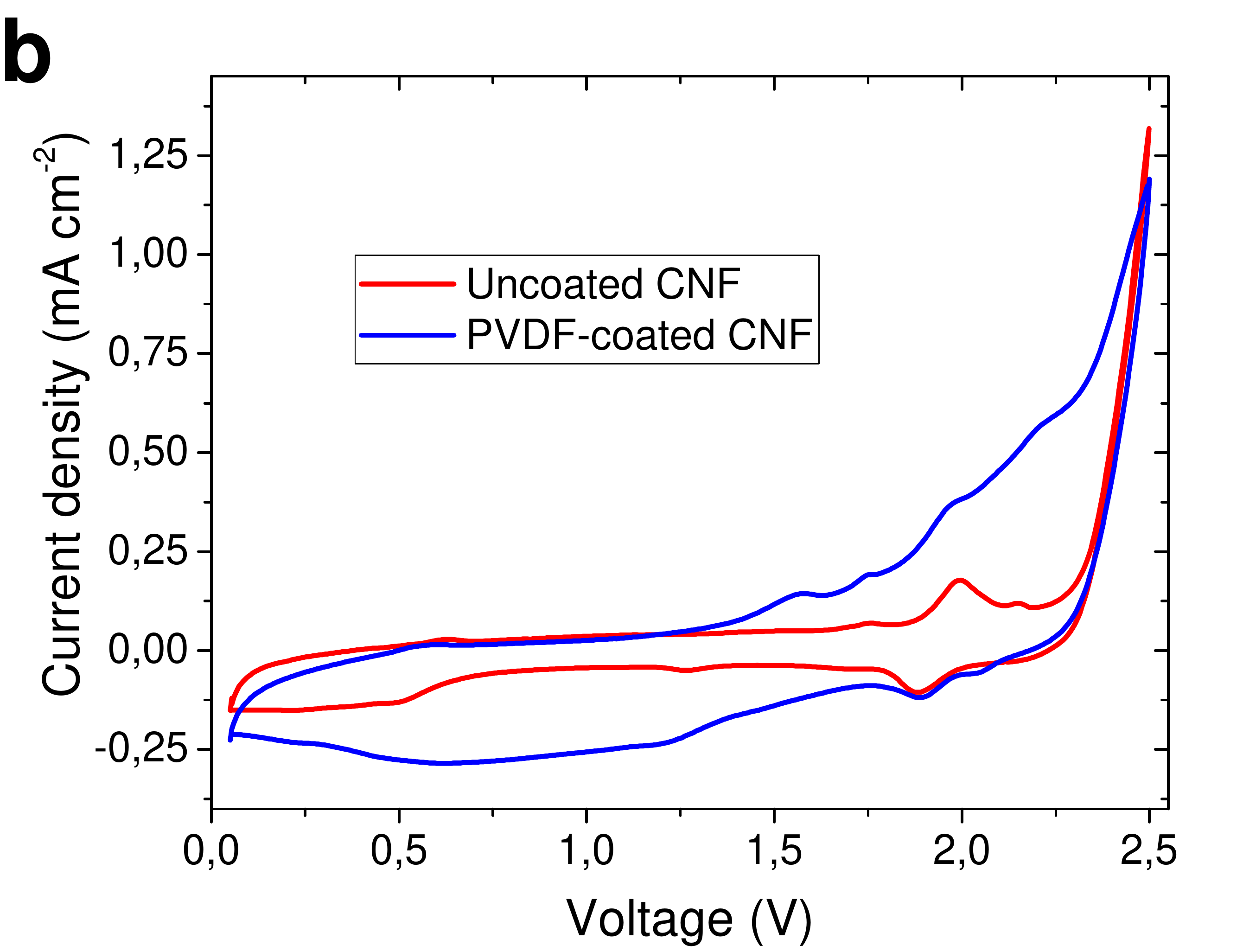}
  \includegraphics[width=0.4\textwidth]{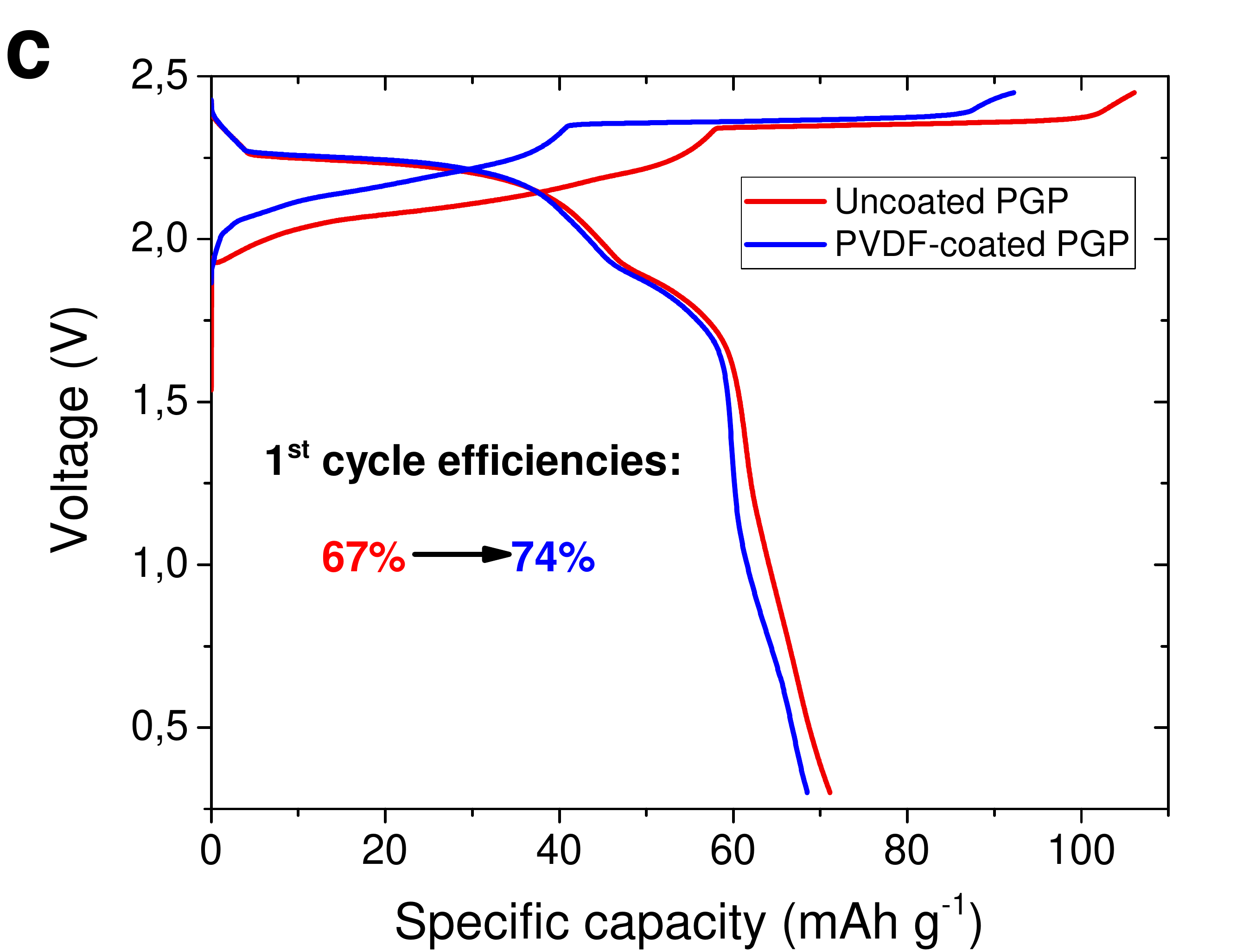}
  \caption{Comparison between the electrochemical performance of Swagelok-type cells built using carbonaceous cathodes with and without PVDF coating: (a) galvanostatic charge-discharge profiles of CNFs (tenth cycle, 100 mA g$^{-1}$), (b) cyclic voltammetries of CNFs (third cycle, 10 mV s$^{-1}$), and (c) comparison of the first galvanostatic charge-discharge cycle (50 mA g$^{-1}$) between uncoated and PVDF-coated PGP.}
  \label{fig:fig9}
\end{figure}

Galvanostatic charge-discharge tests were performed on the materials to test the effect of PVDF coating on CNFs. In order to perform a more appropriate comparison, a specific capacity-based cut-off was used for the charging step (44 mAh g$^{-1}$ for both materials) instead of a more conventionally used voltage-based cut-off. It can be seen from Figure \ref{fig:fig9}a that for the PVDF-coated CNFs, the specific discharge capacity was more than tripled compared to their uncoated counterpart, and a more defined plateau was observed in the discharge curve at about 1.6 V, indicating the presence of a more prominent faradaic process. Furthermore, the charging curve relative to the coated sample does not feature the typical horizontal plateau at low voltages observed in previous experiments, but instead reaches higher voltages within the charging capacity limits of the experiment. This enhanced performance was stable across a significant number of cycles, and at various current rates (Figure S7
).

Cyclic voltammetry also confirmed the trends observed in the galvanostatic charge-discharge tests: it can be seen from Figure \ref{fig:fig9}b that the oxidation peaks between 1.5 and 2.2 V were notably more intense in PVDF-coated CNFs, and a series of new broad reduction peaks also appears at lower voltages. Furthermore, the increase in current at high potentials, associated with the bulk oxidation of the electrolyte, was notably less sharp for PVDF-coated CNFs. The scanning-rate dependency of the voltammetric current (Figure S8
) also confirms that a higher faradaic contribution is present, suggesting a more prominent intercalation mechanism. 

These results therefore suggest that PVDF can indeed effectively shield the surface defects of CNFs and inhibit the formation of an SEI. In addition, the dramatic increase in specific discharge capacity and appearance of plateaus indicate that even materials with a low degree of graphitic character, such as CNFs, could allow a certain degree of intercalation-based energy storage mechanism. However, this was only possible if the surface of the material was coated with an inert film, that can shield any surface defects present, thus inhibiting the unwanted processes of electrolyte decomposition and SEI formation. Further studies would be needed to verify this hypothesis; nevertheless, it is worth considering that the coating process was optimized to not disrupt the nanofibrous structure of CNFs, but only coat individual nanofibers (Figure S9
): the two samples should therefore have comparable surface area, and the performance enhancement should hence not be attributed to a variation in the nanomorphology of the material. Furthermore, a control galvanostatic charge-discharge experiment performed on pure electrosprayed PVDF (Figure S10
) revealed that the polymer had negligible activity as AIB cathode. Therefore, the hypothesis that the performance improvement might be attributed to side reactions involving the PVDF coating was also excluded. Finally, a moderate performance enhancement was also observed when the highly graphitic PGP was coated with PVDF, as demonstrated by the increase of coulombic efficiency of the first cycle in PVDF-coated PGP (Figure \ref{fig:fig9}). This indicates that even in this case, surface defects were responsible for the loss in efficiency during the first cycle. In uncoated PGP, the efficiency was recovered in the second cycle, thanks to the newly-formed SEI layer coating the graphite particles and shielding the surface defects. On the other hand, if PGP was coated with PVDF, then the defects present on the surface of the graphite particles were already protected from direct contact with the electrolyte, resulting in a partially suppressed decomposition of the electrolyte, and a higher efficiency in the first cycle. This observation also puts the role of binder polymers in Al-graphite batteries in a new light, revealing a two-fold purpose: 1) allowing the active material to be processed into a thin film, and 2) suppressing the decomposition of the electrolyte by coating the graphite particles and shielding the surface defects responsible for these unwanted processes. Because of this, the use of binder polymers might be advantageous even in the case of free-standing carbon materials.

\section{Conclusions}
In summary, we have investigated the formation of solid-electrolyte interphases in aluminum-ion batteries featuring carbonaceous cathodes. This process was found to be primarily responsible for the poor coulombic efficiencies observed in the first cycles by recent publications. Using electrospun carbon nanofibers with a varying degree of crystallinity, we have proven that this phenomenon is caused by an early onset of decomposition of the electrolyte, taking place at cell potentials lower than the commonly used 2.45 V cut-off value. Using a combination of characterization techniques, the origin of this process was found to be related to the presence of defect sites and heteroatomic functional groups on the surface of the material, which could likely form Lewis adducts and react with the chloroaluminate species to generate insoluble compounds, leading to the SEI film. This effect was even more prominent with high surface areas cathode materials, as the density of reactive functional groups in the materials increases accordingly. Further confirmation of these hypotheses was found in control experiments performed by coating the cathode materials with the inert binder polymer PVDF, which can effectively shield the reactive species from direct contact with the electrolyte, inhibiting SEI formation. This modification led to improved electrochemical performance in the cathode materials: PVDF-coated CNFs obtained a significant boost in specific discharge capacity, suggesting the possibility of an intercalation-based energy storage mechanism, and the coulombic efficiency relative to the first cycle of PGP was increased.

The results presented in this work bring some clarity to some of the ambiguities and discrepancies present in current AlCl3-graphite battery literature, and therefore will hopefully be useful in better understanding the processes taking place in the system, and ultimately help create better performing devices.

\section{Experimental section}
Polyacrylonitrile (average M$_w$: 150,000), N,N’-dimethylformamide, N,N’-dimethylacetamide aluminum trichloride, 1-ethyl-3-methylimidazolium chloride, poly(vinylidene fluoride) (average M$_w$: 534,000) and cobalt acetate tetrahydrate were purchased from Sigma-Aldrich. Pyrolytic graphite paper was purchased from MTI corporation. Al Foil was purchased from Goodfellow. All chemicals and materials were used without any additional treatment, unless differently specified.

SEM micrographs and EDXS data were acquired with a JEOL-JSM6500F field emission scanning electron microscope, using an accelerating voltage of 20 kV. Electrosprayed PVDF was sputter-coated with a 5 nm layer of platinum to enhance image quality. All the remaining samples were imaged without the use of any coating. For ex-situ experiments, the cycled devices were disassembled inside the glovebox, and the cathodes were washed with ethanol to remove any residual electrolyte. 

X-ray diffractograms were acquired using a PANalytical X'pert PRO XRD diffractometer operating in reflection mode, using a Cu K$\alpha$ X-ray source.

TEM micrographs were obtained and STEM mapping were performed on a JEOL 2100 operating at an accelerating voltage of 200 kV. Samples were prepared by sonication in Ethanol and dispersion onto a lacey carbon grid.

BET surface area measurements were performed using a Micrometrics FlowSorb II 2300 instrument at a temperature of 77 K using a 30\% N2 in He mixture.

Raman spectra were acquired using a LabRAM HR800 Raman spectrometer (Horiba Jobin-Yvon), using an excitation wavelength of 514.5 nm, a grating of 1800 gr/mm, and objective magnification of 100x. The acquisition time was approximately 10 seconds for each spectrum. 

XPS spectra were acquired using a Kratos Axis UltraDLD X-ray Photoelectron Spectrometer (Kratos Analytical, Manchester, UK) equipped with a hemispherical electron energy analyzer. Spectra were excited using monochromatic Al K$\alpha$ X-rays (1486.69 eV) with the X-ray source operating at 150W. This instrument illuminates a large area on the surface and then using hybrid magnetic and electrostatic lenses collects photoelectrons from a desired location on the surface. In this case the analysis area was a 300 by 700 $\mu$m spot obtained using the hybrid magnetic and electrostatic lens and the slot aperture. Measurements were carried out in a normal emission geometry. Core level scans were collected with a pass energy of 20 eV. The analysis chamber was at pressures in the 10$^{-9}$ torr range throughout the data collection. Data analysis was performed using the CasaXPS software.

Cyclic voltammetry tests were performed using a Metrohm Autolab PGSTAT128N potentiostat, operating in a 2-electrode configuration: the cathode terminals of the devices were connected to the working electrode, and the anode terminals were connected to the counter and reference electrodes. The current densities are calculated relatively to the macroscopic surface area of the working electrode. 

Galvanostatic charge-discharge experiments were performed using a NEWARE BTS CT-4008-5V10mA-164 (MTI Corp.) battery analyzer system. Unless differently specified, the charge-discharge experiments were run between 0.3 and 2.45 V. The reported specific capacities are relative to the mass of carbonaceous material in the cathode. 

\subsection{Fabrication of CNFs}
Precursor solutions were prepared by adding 0.8 mL of to 10 mL of DMF under magnetic stirring. For enhanced CNFs, 0.8 g of cobalt acetate tetrahydrate were also added to the solution. The solutions were then loaded into plastic syringes with a stainless steel, 18-gauge, blunt tip needle. The electrospinning process was then performed, using a positive voltage of 15 kV, a flow rate of 1 mL/hour, and a tip-to-collector distance of 13 cm. A rotating drum collector was used in the process to optimize the form factor of the nanofiber mats.
The resulting fibers were first heat-treated in air using a muffle furnace with the following heating program: heating at 5 \degree ree  C/minute until 220 \degree  C, holding the temperature for 2 hours, and cooling naturally to room temperature. Finally, the fibers were carbonized under a nitrogen atmosphere using a tube furnace with the following heating program: heating at 4 \degree  C/minute until 200 \degree  C, hold temperature for 20 minutes, heating at 10 \degree  C/minute until 600 \degree  C, hold temperature for 20 minutes, heating at 5 \degree  C/minute until 1400 \degree  C, hold temperature for 1 hour, cooling at 5 \degree  C/minute until 200 \degree  C, and finally cooling naturally to room temperature. For enhanced CNFs, the carbonized mats were soaked repeatedly in concentrated aqueous HCl until no blue color (relative to Co$^{2+}$ complexes) could be observed.  

\subsection{Fabrication of carbon film}
A precursor solution was prepared by adding 0.8 mL of PAN to 10 mL of DMF under magnetic stirring. Roughly 3 mL of this solution were then poured into a Petri dish, and the solvent was evaporated in air over 2 hours. The resulting PAN film was first stabilized in air using a muffle furnace with the following heating program: heating at 5 \degree C/minute until 220 \degree C, holding the temperature for 2 hours, and cooling naturally to room temperature. Finally, the material was carbonized under a nitrogen atmosphere using a tube furnace with the following heating program: heating at 4 \degree C/minute until 200 \degree C, hold temperature for 20 minutes, heating at 10 \degree C/minute until 600 \degree C, hold temperature for 20 minutes, heating at 5 \degree C/minute until 1400 \degree C, hold temperature for 1 hour, cooling at 5 \degree C/minute until 200 \degree C, and finally cooling naturally to room temperature.

\subsection{Fabrication of electrosprayed PVDF}
A precursor solution was prepared by adding 1 g of PVDF (average M$_w$: 534,000) to 10 mL of N,N’-dimethylacetamide and stirring at 70 \degree C for 2 hours. After complete dissolution of the polymer, the solution was cooled to room temperature, and loaded into a plastic syringe with a stainless steel, 18-gauge, blunt tip needle. The solution was electrosprayed using a positive voltage of 15 kV, a flow rate of 1 mL/hour, and a tip-to-collector distance of 13 cm. A rotating drum collector was used in the process to optimize the form factor of the mats. After approximately 5 mL of solution were electrosprayed, the process was interrupted, and the mat was removed from the current collector. 

\subsection{Dip-coating of CNF and PGP}
A poly(vinylidene fluoride) solution was prepared by adding 0.4 g of PVDF (average M$_w$: 534,000) to 10 mL of N,N’-dimethylacetamide and stirring at 70 \degree C for 2 hours. After complete dissolution of the polymer, the solution was cooled to room temperature. CNF or PGP mats were manually dipped into the solution and extracted using a pair of reverse action tweezers, and immediately transferred to a vacuum oven to be dried at 150 \degree C for 1 hour. 

\subsection{Electrolyte preparation}
[EMIm]Cl was baked at 100 \degree C in a vacuum oven for 48 hours to remove its water content, and immediately brought inside a nitrogen-filled glovebox with the O$_2$ and H$_2$O levels kept below 1 ppm. The electrolyte was prepared by gradually adding 1.3 equivalents of anhydrous AlCl$_3$ to 1 equivalent [EMIm]Cl inside the glovebox. An exothermic reaction takes place, after which a brown, slightly viscous liquid is formed. 

\subsection{Device construction}
CNF mats, PGP, carbon films, and electrosprayed PVDF films were cut into 11 mm diameter discs to be used as cathode in the devices. Prototype batteries were then assembled using custom-built Swagelok-type cells consisting of a polyether ether ketone “straight union” pipe fitting, with an inner diameter of 12 mm, and two molybdenum rods with a diameter of 12 mm as terminals; two rubber O-rings, placed at each end of the fitting, ensured that the inner volume of the device was sealed from the external environment (Figure S11 
). All the components of the device were first baked at 100 \degree C in a vacuum oven for at least 2 hours to remove any residual water, then immediately transferred inside a nitrogen-filled glovebox with the O$_2$ and H$_2$O levels kept below 1 ppm. 11 mm discs of either CNF, PGP, carbon film, or electrosprayed PVDF mats were used as cathodes. 11-mm diameter aluminum foil discs were used as the anode. Glass microfiber (GF/D) discs with 12 mm diameters, soaked in the electrolyte (approximately 400 $\mu$L), were used as separators. The devices were then wrapped with Parafilm as an additional moisture barrier, and taken outside the glovebox for electrochemical testing.

\section*{Acknowledgements}
We would like to Acknowledge Dr.\ Johan Grand for his help with the acquisition of Raman Spectra, and David Flynn for assisting with SEM and TEM measurements. Thanks also to Dr.\ Colin Doyle from the University of Auckland for his help with XPS experiments.

\bibliographystyle{unsrt}  
\bibliography{refs}


\end{document}


\beginsupplement
\cleardoublepage

\section*{Solid-electrolyte interphases (SEI) in nonaqueous aluminum-ion batteries}
\textit{Nicolò Canever, Fraser R.\ Hughson, and Thomas Nann}
\subsection*{Supporting Information}


\begin{figure} [h!]
    \centering
    \includegraphics[width=0.45\linewidth]{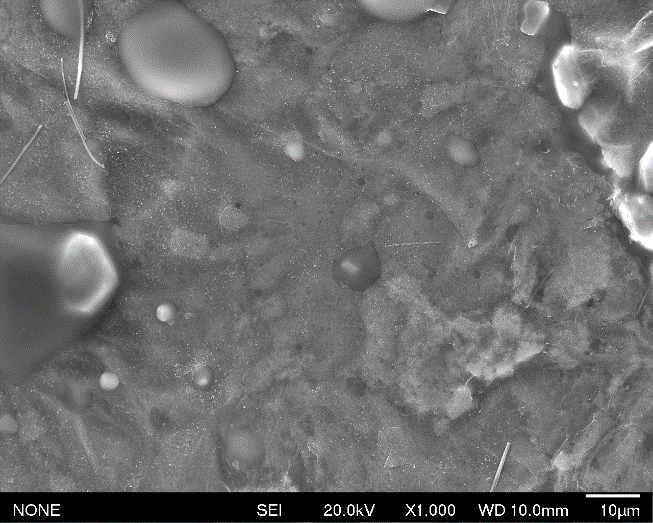}
    \includegraphics[width=0.45\linewidth]{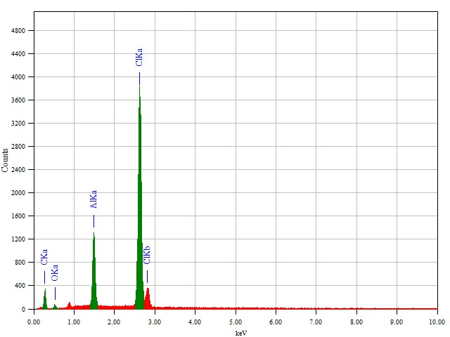}
    \caption{SEM micrograph (left) and relative EDXS spectrum (right) of pyrolytic graphite paper, after a typical galvanostatic charge-discharge test (20 cycles at a current rate of 50 mA g$^{-1}$).}
    \label{fig:figS1}
\end{figure}

\begin{figure} [h!]
    \centering
    \includegraphics[width=0.6\linewidth]{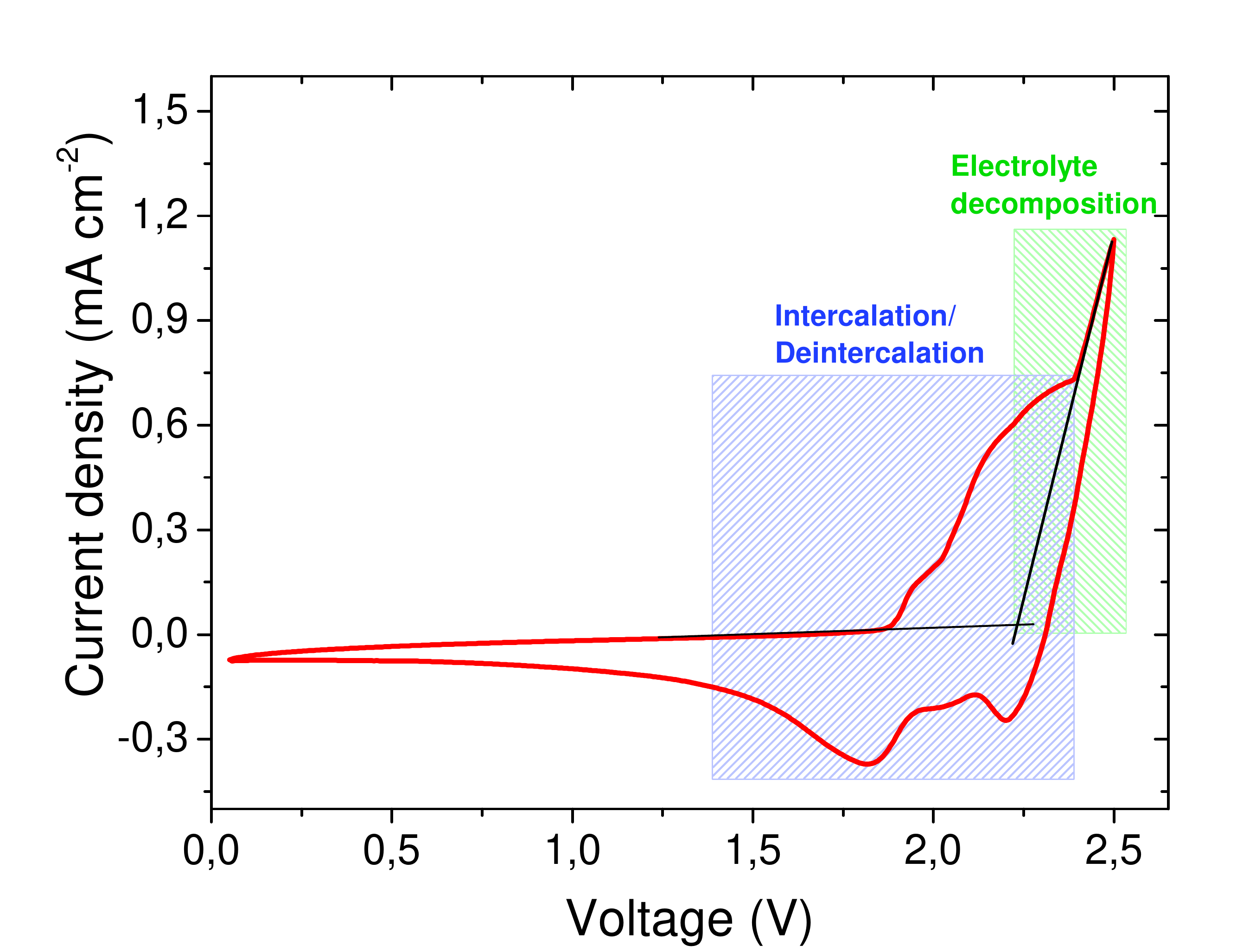}

    \caption{Cyclic voltammogram (third cycle, 10 mV s$^{-1}$) of a Swagelok-type cell built using pyrolytic graphite paper as cathode.}
    \label{fig:figS2}
\end{figure}

\begin{figure} [h!]
    \centering
    \includegraphics[width=0.6\linewidth]{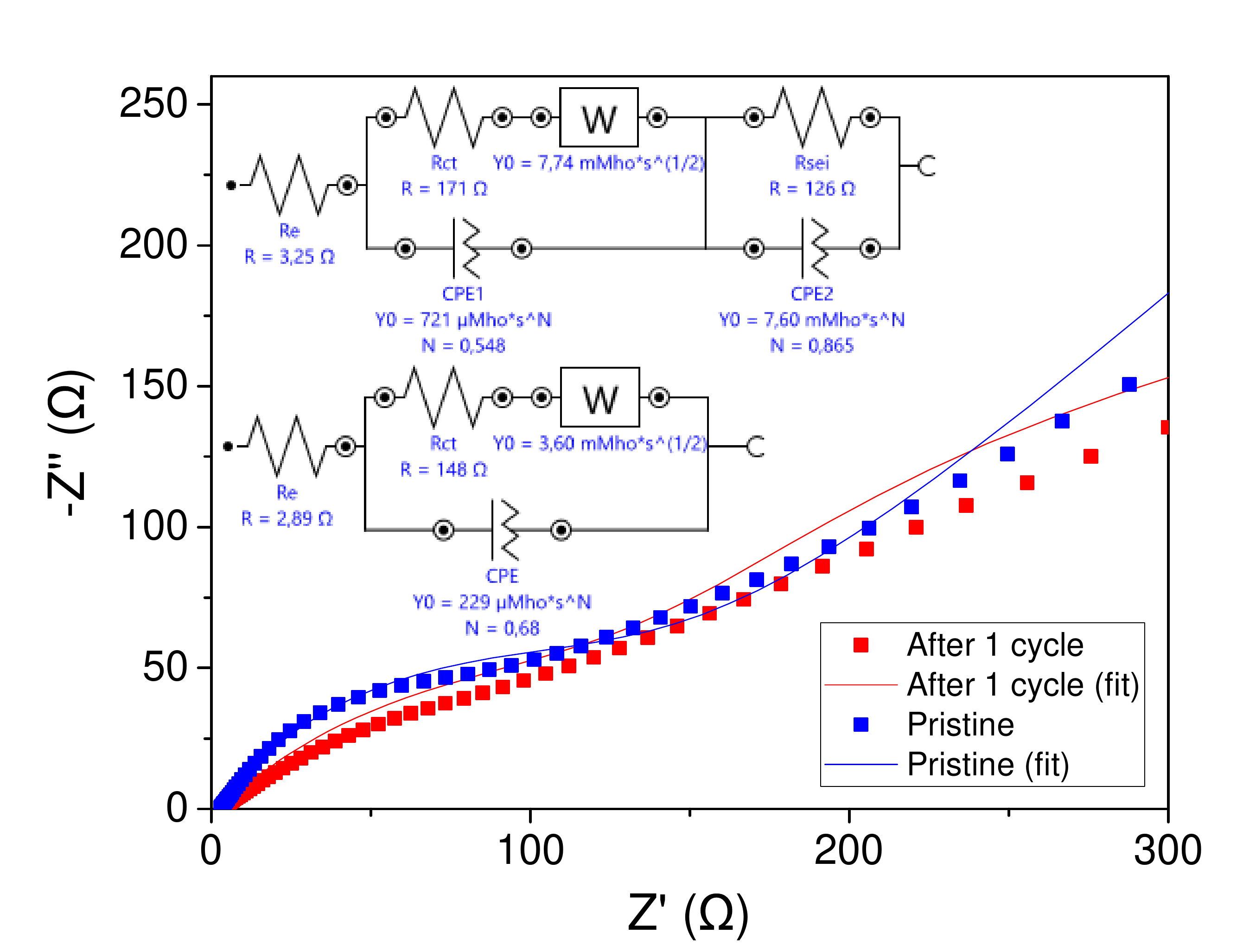}

    \caption{Nyquist plots of a Swagelok-type cell built using pyrolytic graphite paper as cathode before and after the first charge discharge cycle at 50 mA g$^{-1}$. It can be seen from the graph that the high-frequency region of the plots changes significantly after the first charge-discharge cycle. This is consistent with our interpretation of the formation of an SEI. A simple Randles circuit (bottom schematic) was used for fitting the plot relative to the pristine cathode. On the other hand, for the cycled cathode, a resistor-constant phase element parallel pair is added in series to the rest of the circuit (top schematic), which is indicative of a poorly-conducting film coating the electrode.\cite{barsoukov_impedance_2018}
}
    \label{fig:figS3}
\end{figure}

\begin{figure} [h!]
    \centering
    \includegraphics[width=0.6\linewidth]{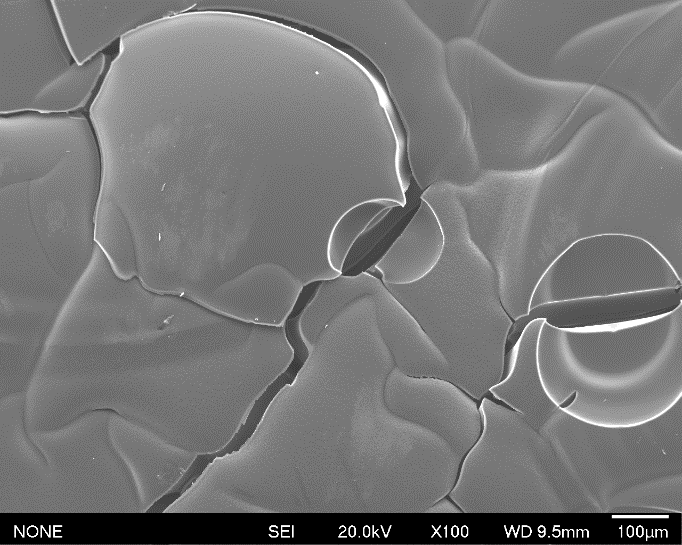}

    \caption{SEM micrograph of the PAN-derived carbon film.}
    \label{fig:figS4}
\end{figure}

\begin{figure} [h!]
    \centering
    \includegraphics[width=0.45\linewidth]{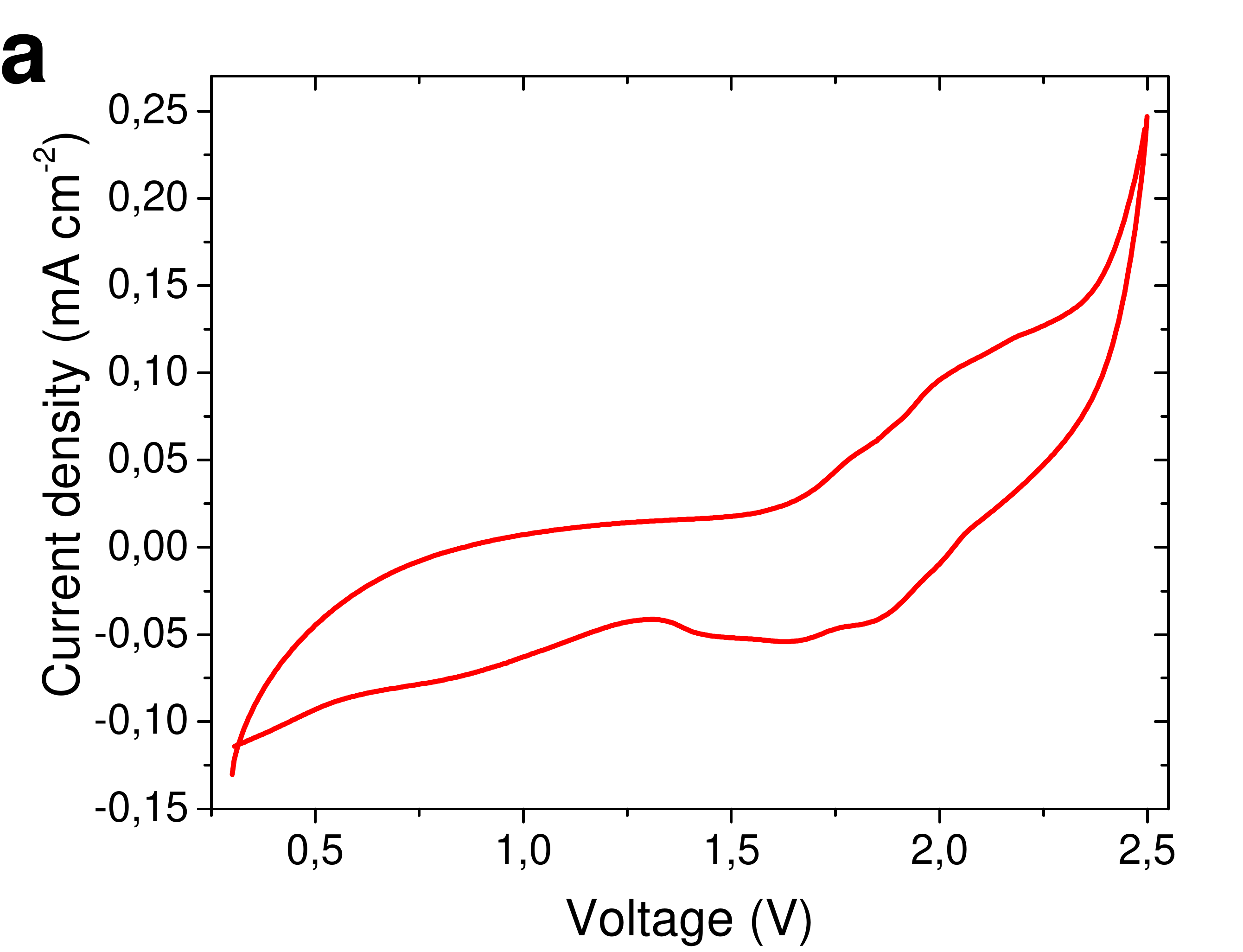}
    \includegraphics[width=0.45\linewidth]{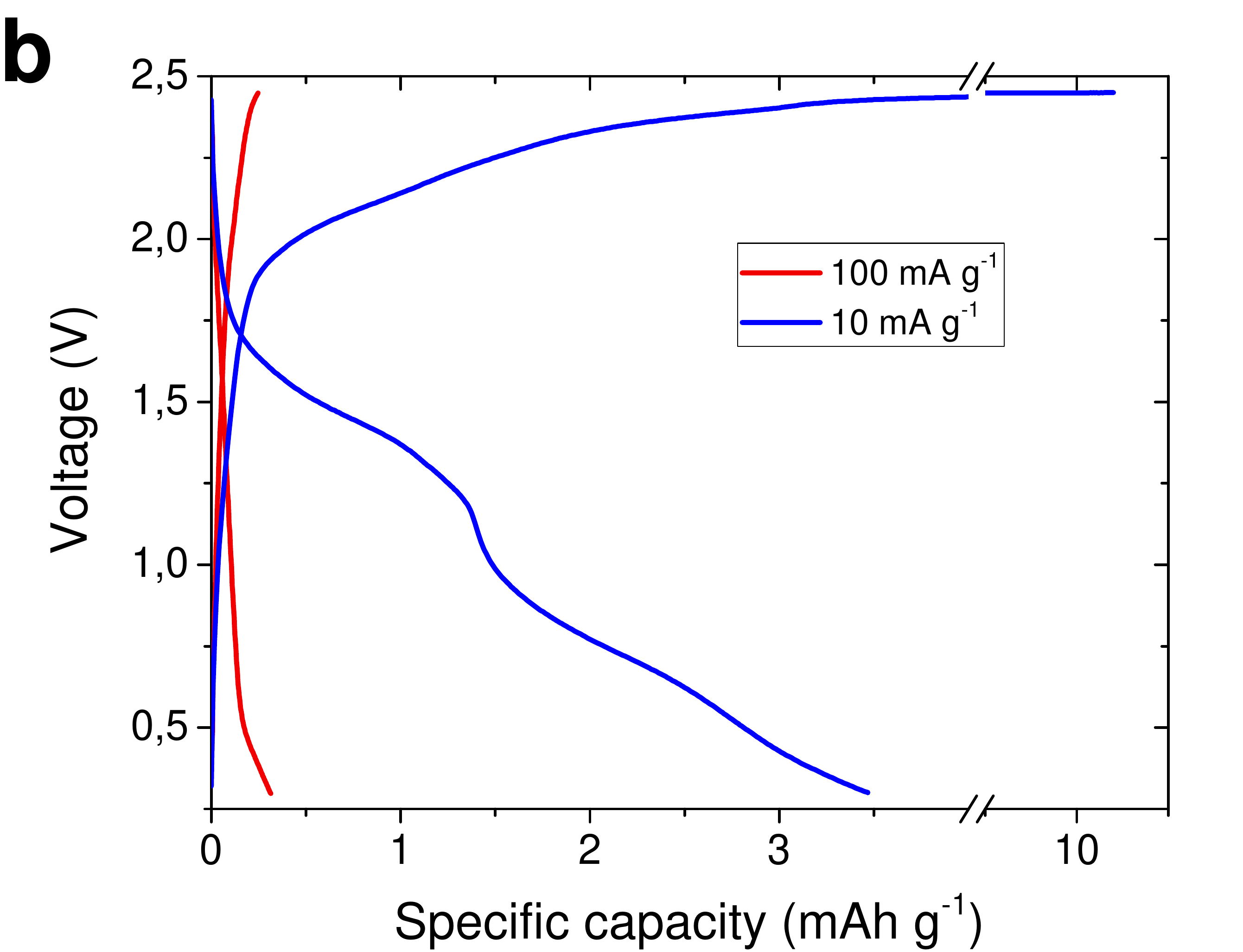}

    \caption{(a) Cyclic voltammetry (third cycle, 10 mV s$^{-1}$) and (b) Galvanostatic charge-discharge profiles (tenth cycle) of a Swagelok-type cell built using the PAN-derived carbon film.}
    \label{fig:figS5}
\end{figure}

\begin{figure} [h!]
    \centering
    \includegraphics[width=0.45\linewidth]{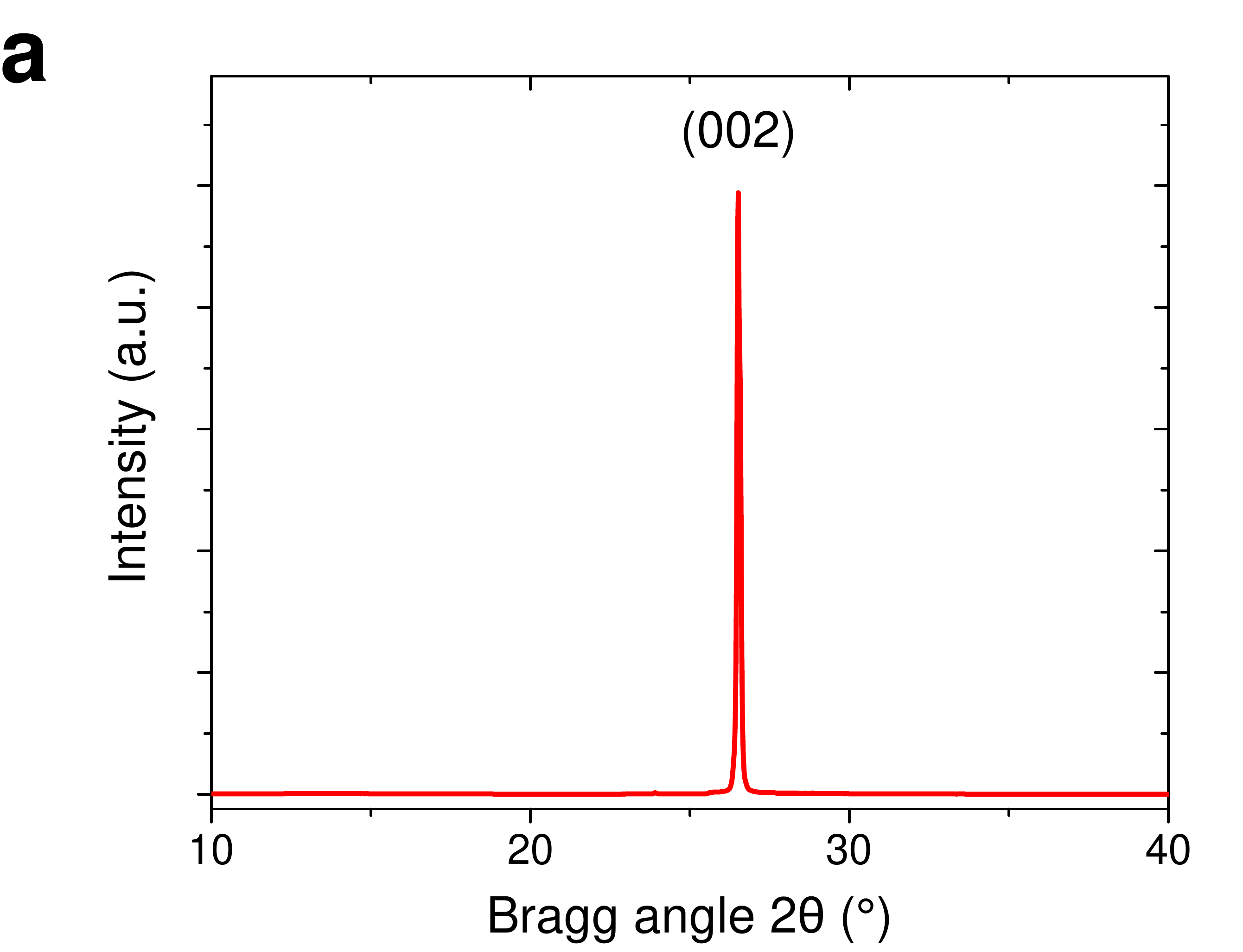}
    \includegraphics[width=0.45\linewidth]{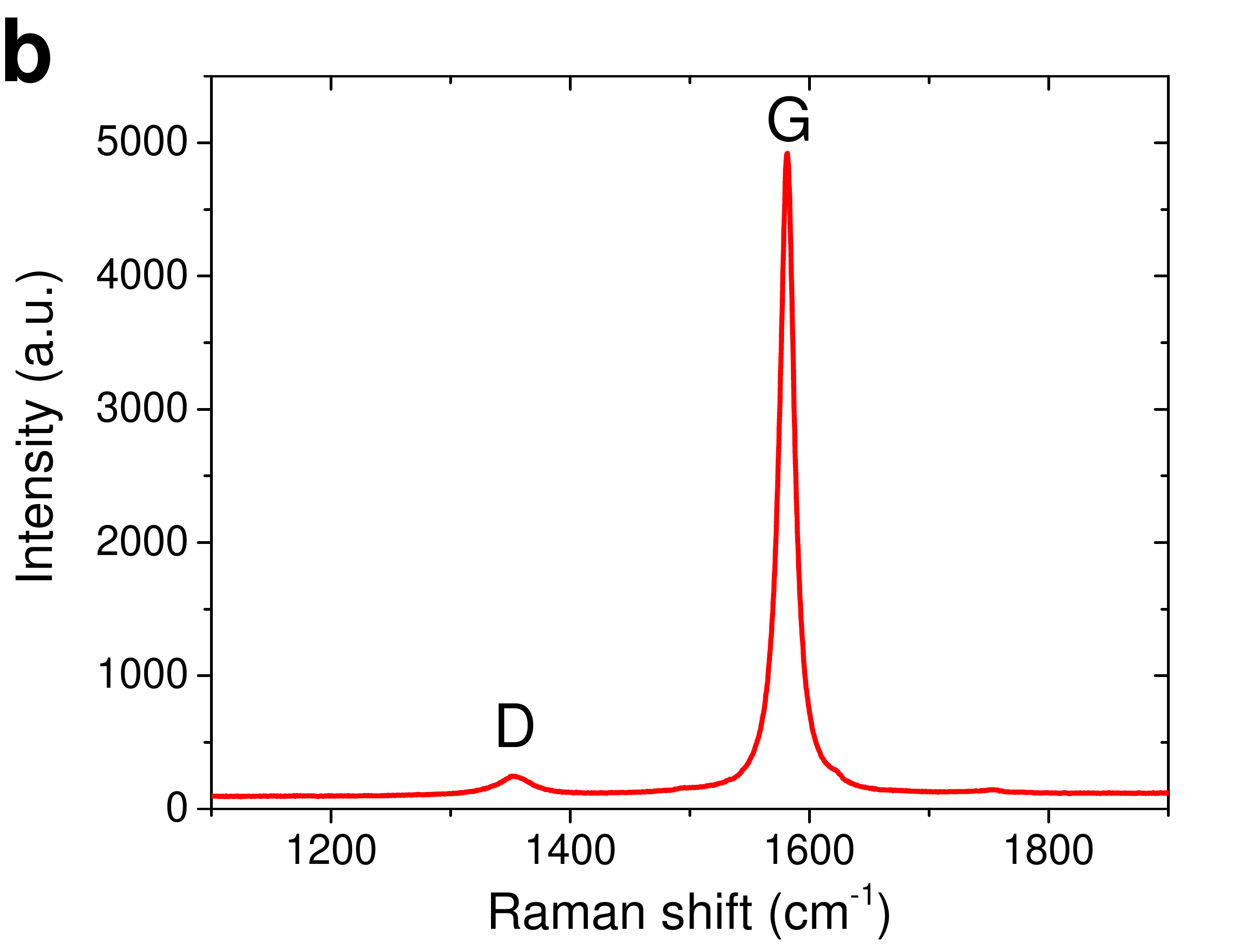}

    \caption{X-ray diffractogram (a) and Raman spectrum (b) of pyrolytic graphite paper.}
    \label{fig:figS6}
\end{figure}

\begin{figure} [h!]
    \centering
    \includegraphics[width=0.45\linewidth]{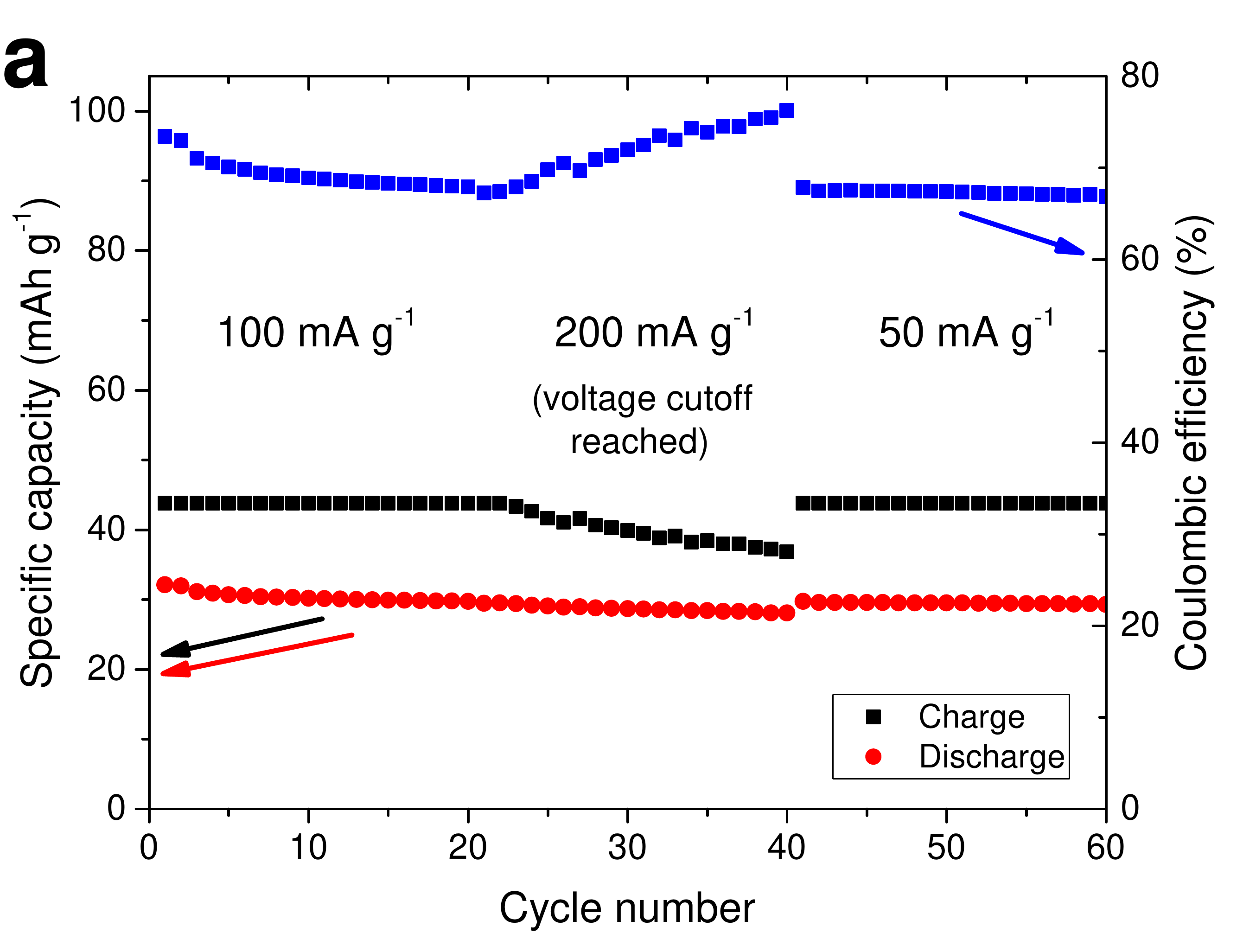}
    \includegraphics[width=0.45\linewidth]{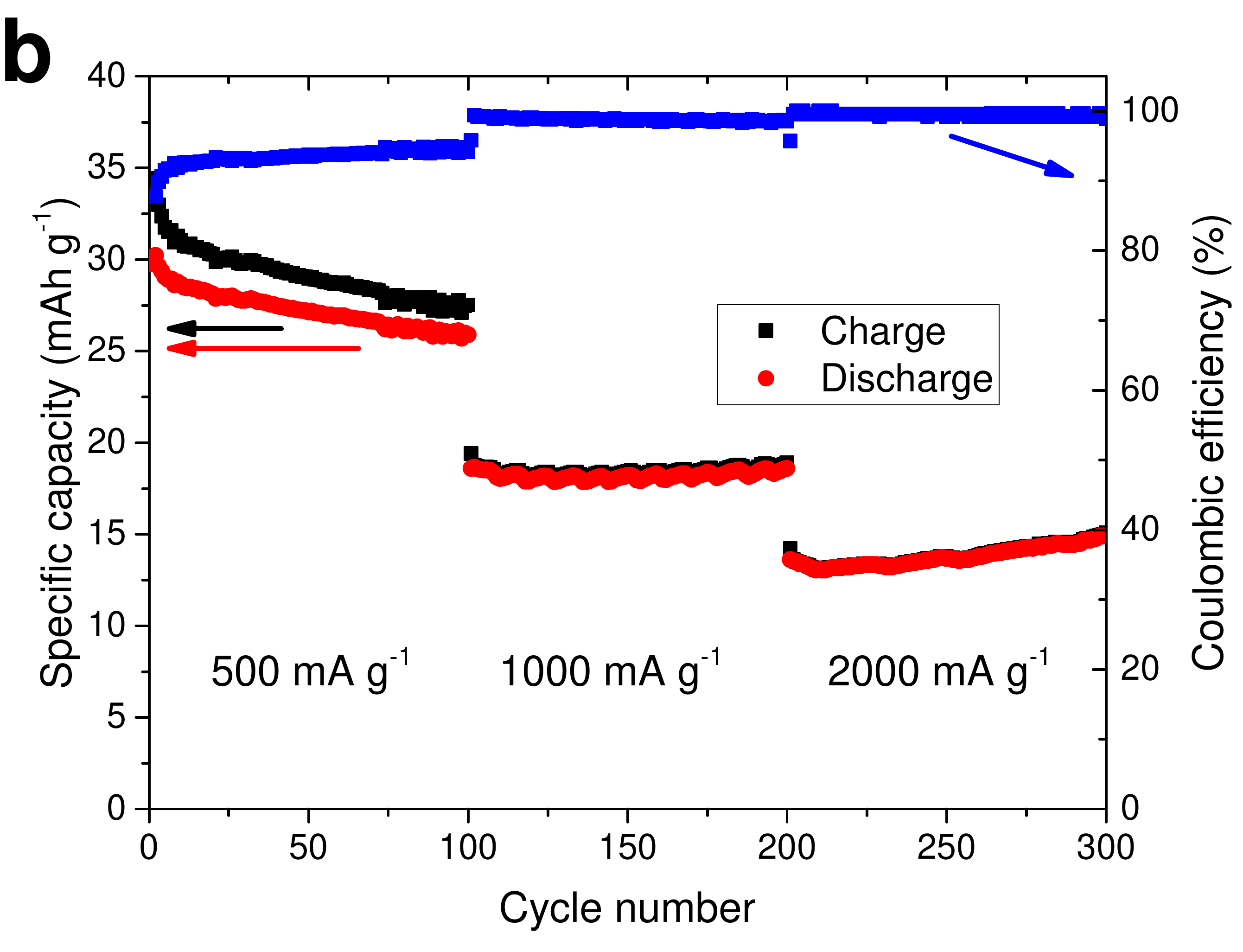}

    \caption{Specific capacities and coulombic efficiencies relative to typical galvanostatic charge-discharge cycles performed on Swagelok-type cells using PVDF-coated CNF as cathodes, at various current rates. A specific charging capacity cutoff of 44 mAh g$^{-1}$ was used in combination with the upper voltage cutoff of 2.45 V in the experiments.}
    \label{fig:figS7}
\end{figure}

\begin{figure} [h!]
    \centering
    \includegraphics[width=0.45\linewidth]{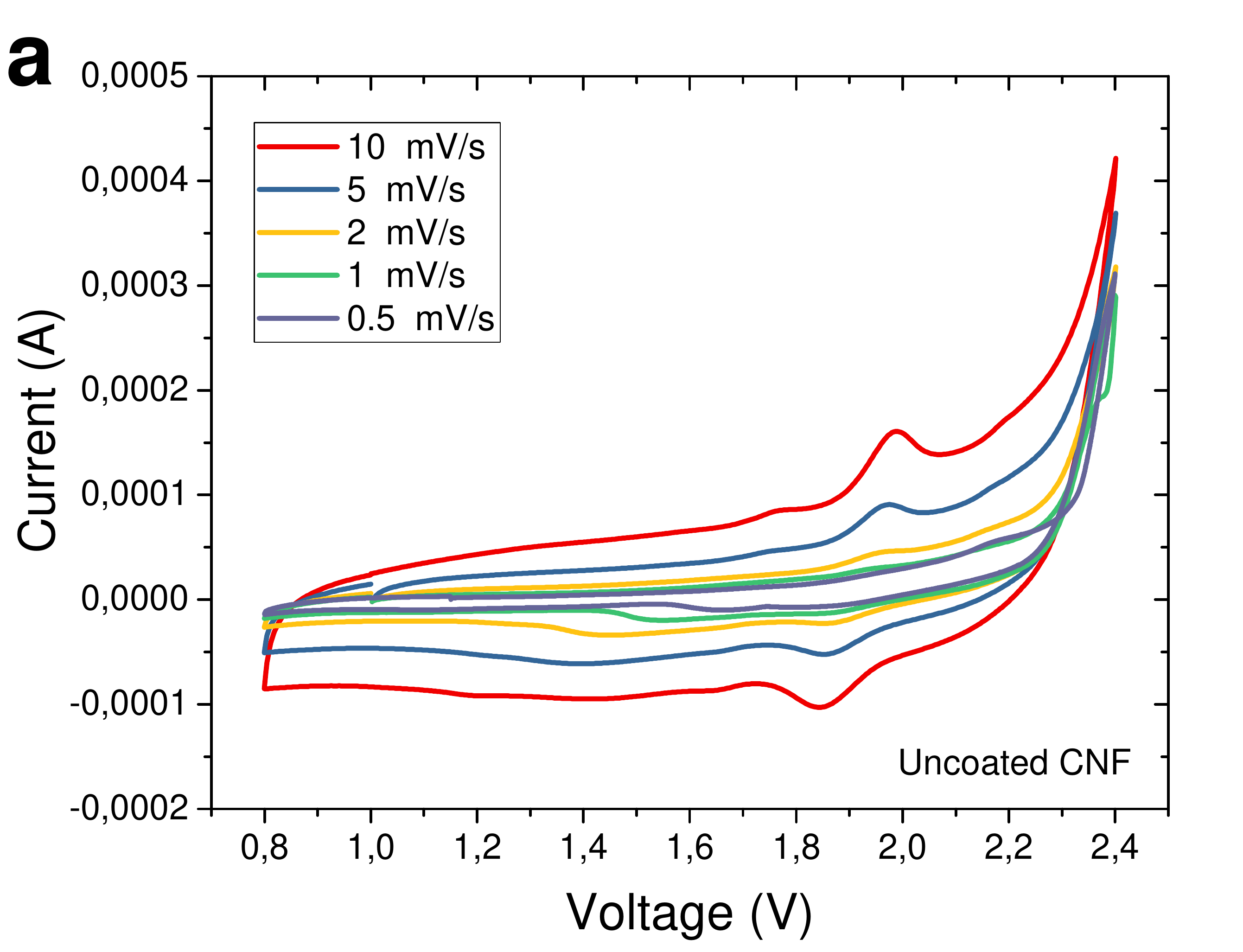}
    \includegraphics[width=0.45\linewidth]{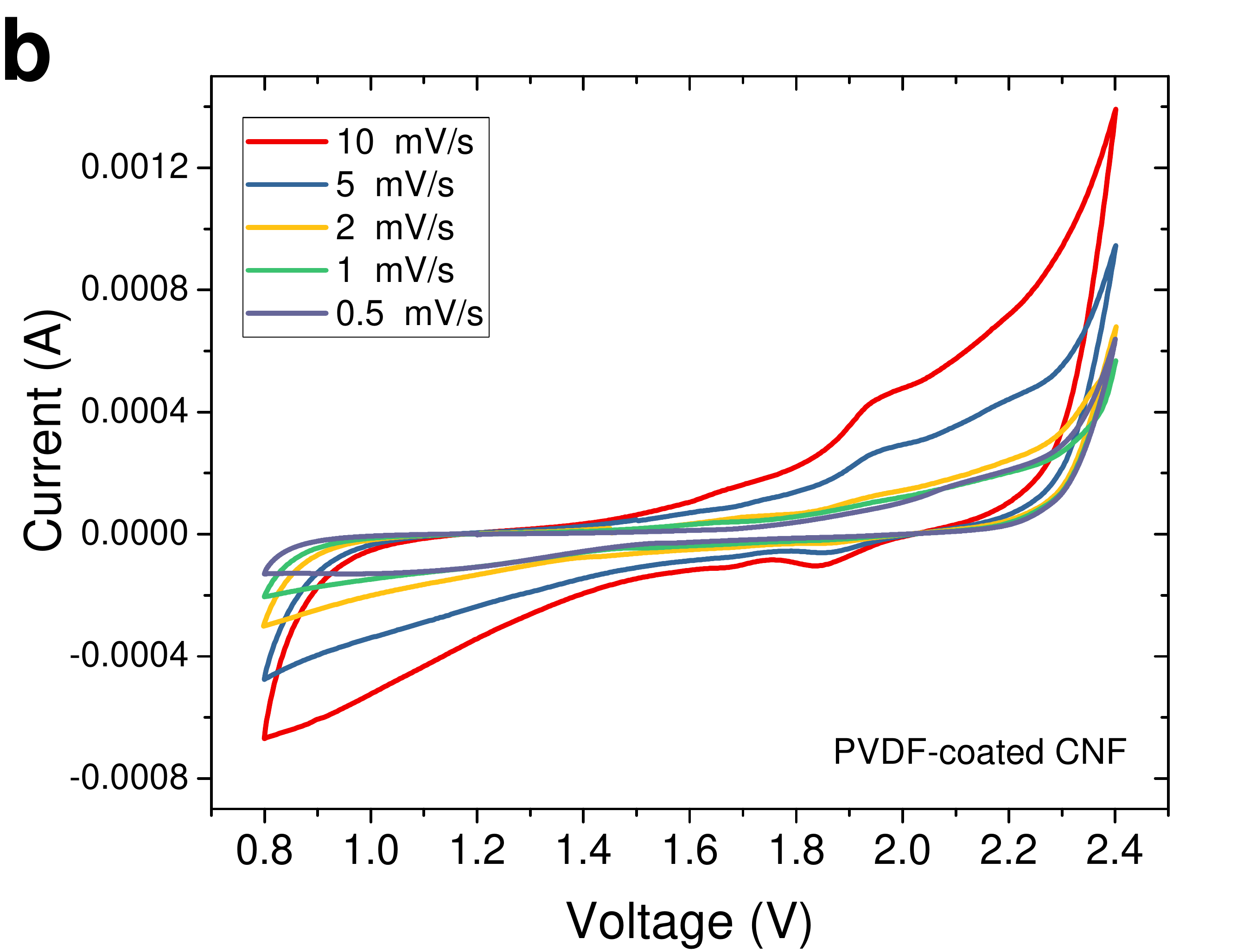}
    \includegraphics[width=0.45\linewidth]{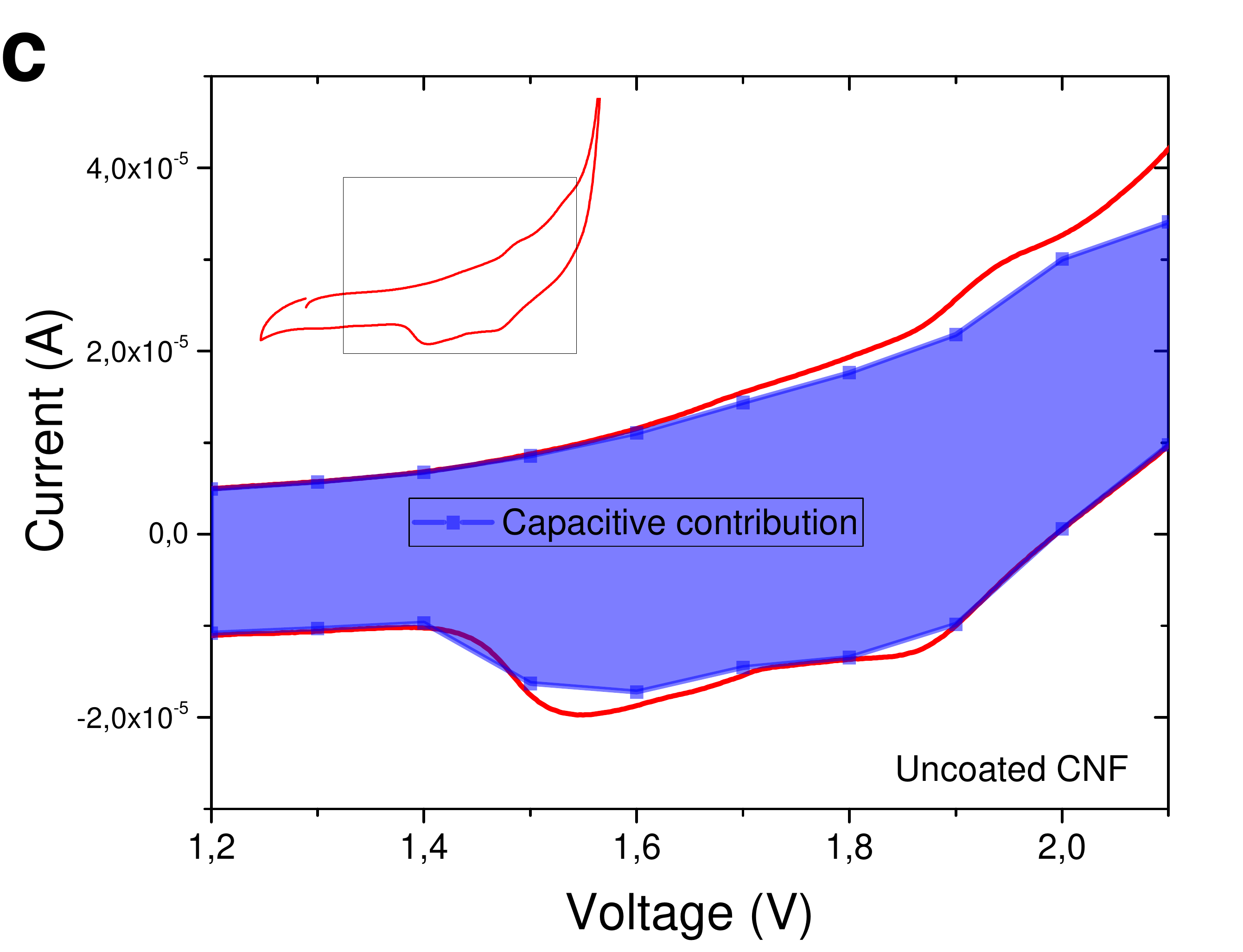}
    \includegraphics[width=0.45\linewidth]{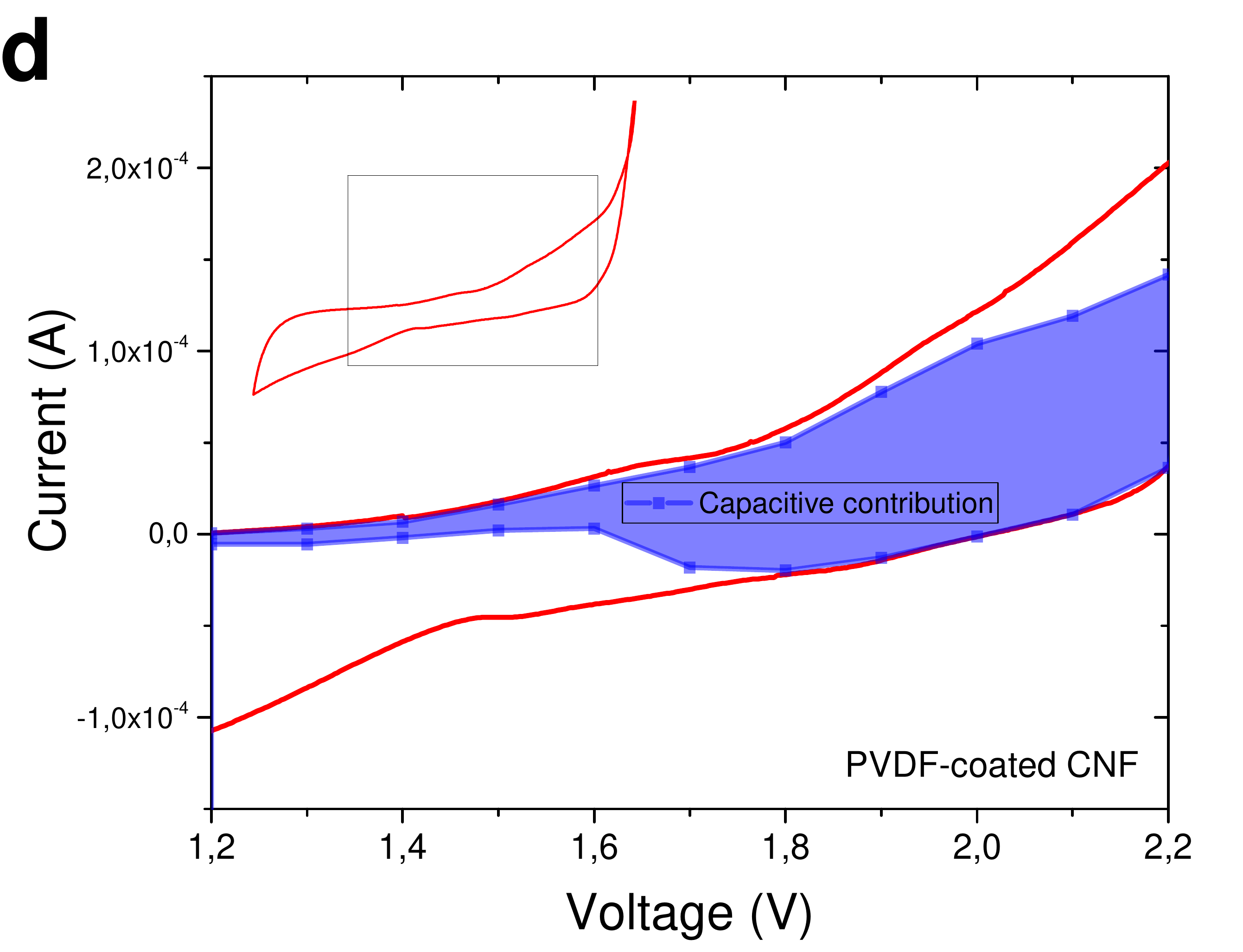}

    \caption{Top: cyclic voltammograms at multiple scan speed of Swagelok-type cells using uncoated (a) and PVDF-coated CNF (b) as cathodes. Bottom: Capacitive current contribution to the voltammetric current for uncoated (c) and PVDF-coated CNF (d), obtained from the above cyclic voltammetry data at different scan speeds, using the technique proposed by Wang et al.\cite{wang_pseudocapacitive_2007}}
    \label{fig:figS8}
\end{figure}

\begin{figure} [h!]
    \centering
    \includegraphics[width=0.6\linewidth]{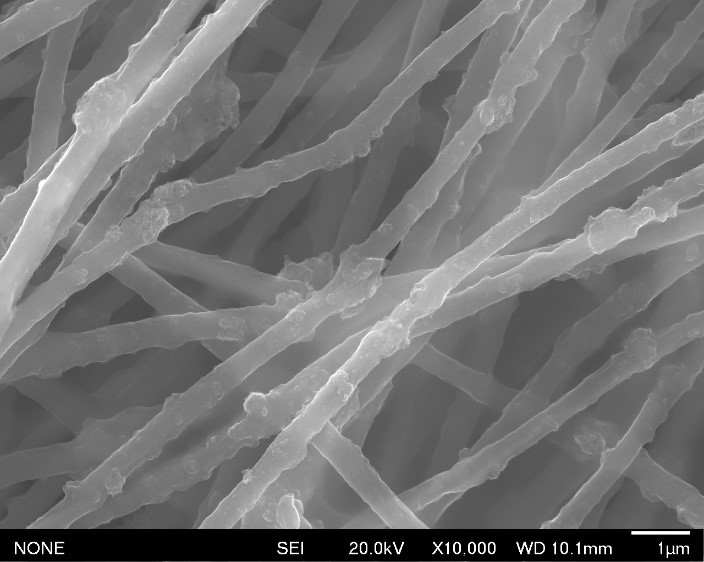}

    \caption{SEM micrograph of PVDF-coated CNF (non-catalyzed).}
    \label{fig:figS9}
\end{figure}

\begin{figure} [h!]
    \centering
    \includegraphics[width=0.6\linewidth]{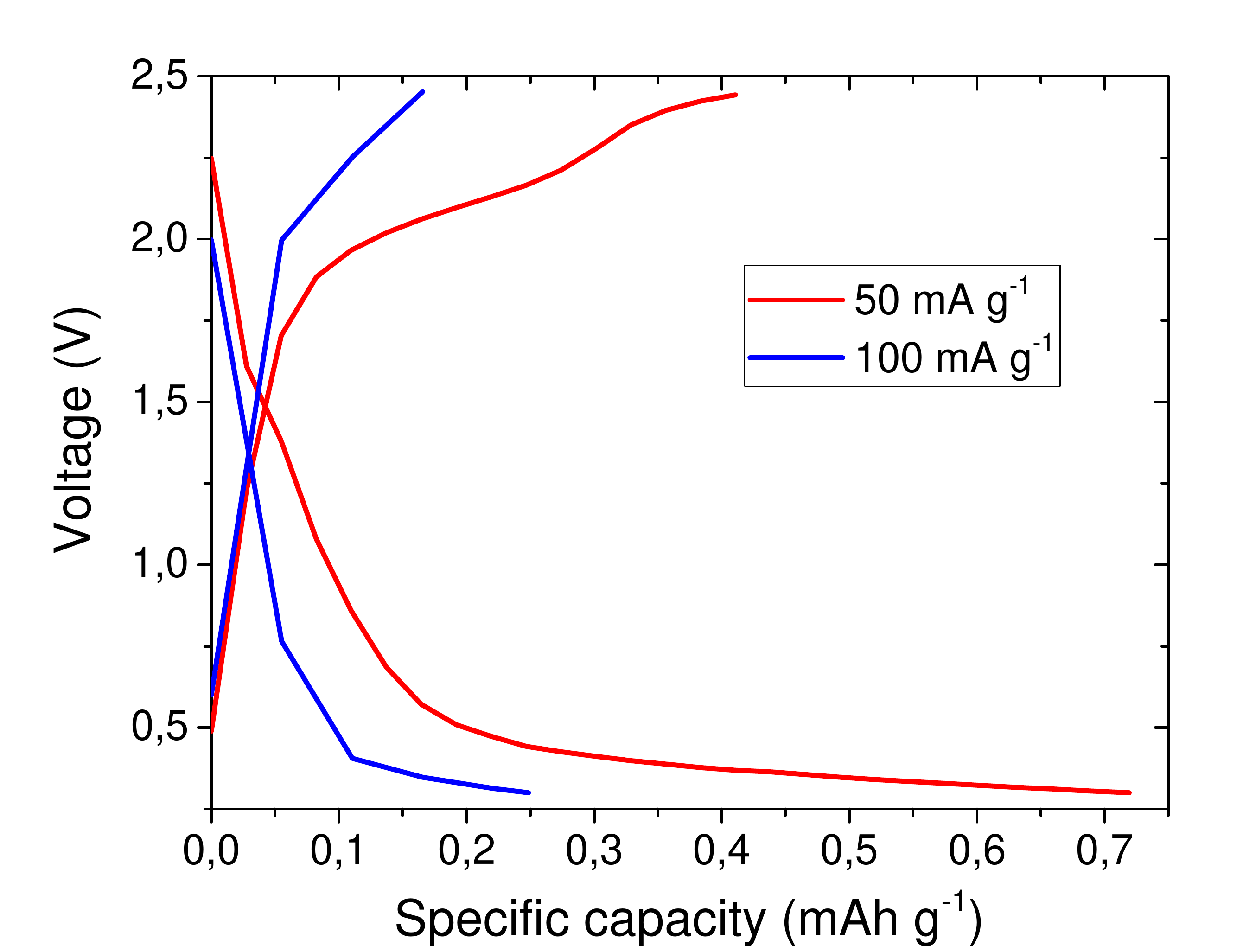}

    \caption{Galvanostatic profiles (tenth cycle) of Swagelok-type cells built using electrosprayed PVDF as cathode.}
    \label{fig:figS10}
\end{figure}

\begin{figure} [h!]
    \centering
    \hspace{2cm}
    \includegraphics[width=0.35\linewidth]{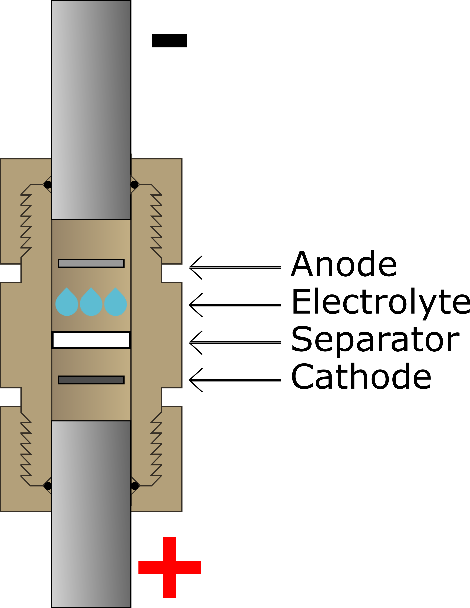}

    \caption{Schematic representation of the custom-built Swagelok-type cell used in this work.}
    \label{fig:figS11}
\end{figure}

\begin{figure} [h!]
    \centering

    \includegraphics[width=\linewidth]{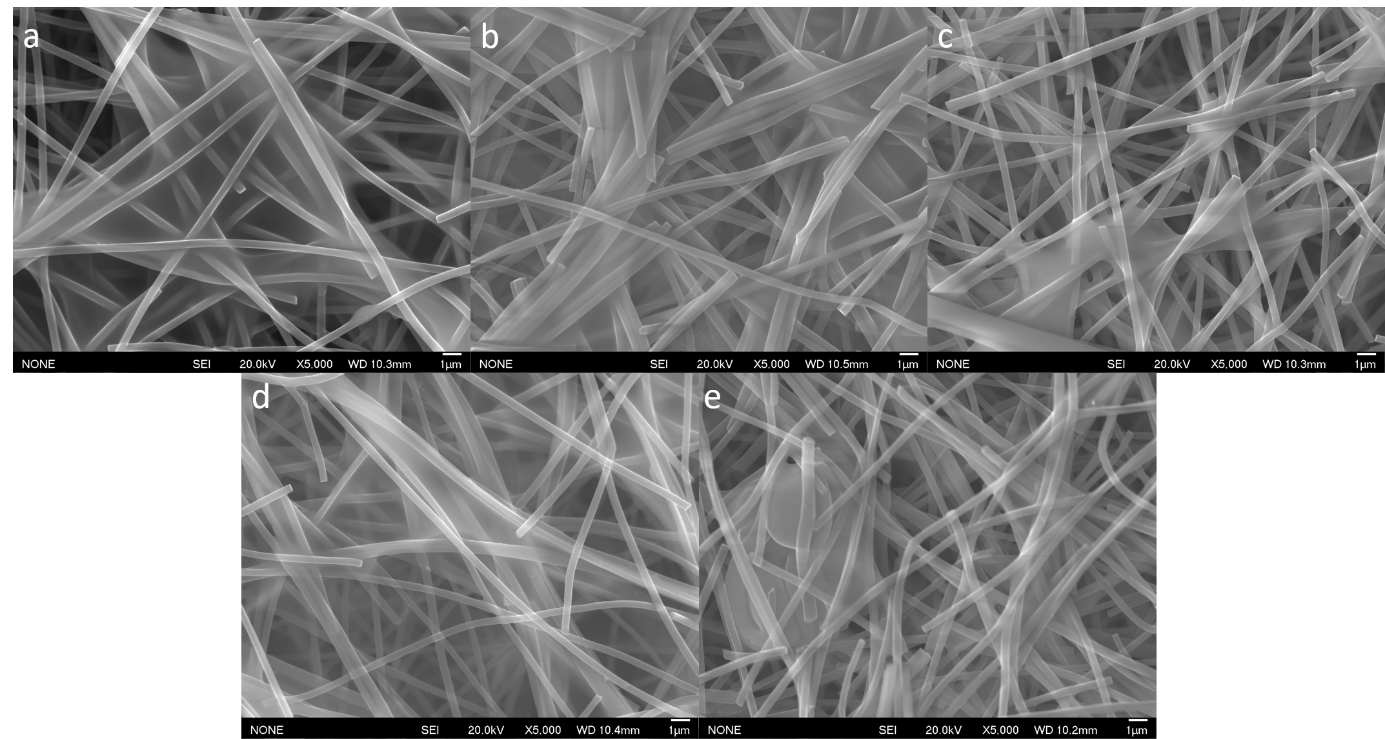}

    \caption{SEM images of cycled cathodes that have been washed by submerging samples in a vial of solvent and agitating: (a) control - no washing, (b) Washed with cyclohexane, (c) washed with N,N'-dimethylformamide, (d) washed with wthanol (e) washed with water. Visually the water had the most effect, removing a lot of the webbing in the sample but some remained after washing. Other solvents had no effect. Both webbing and fibers had near identical elemental compositions and were not affected by washing. }
    \label{fig:figS12}
\end{figure}

\clearpage

\bibliographystyle{ieeetr}
\bibliography{refs}
